\documentclass[prd,twocolumn,nofootinbib]{revtex4-1}
\usepackage{graphicx}
\usepackage{bm}
\usepackage[colorlinks=true]{hyperref}
\usepackage{float}
\usepackage{amsmath}
\usepackage{amssymb}
\usepackage{pifont}
\usepackage{gensymb}
\usepackage{multirow}
\usepackage[dvipsnames]{xcolor}
\usepackage[utf8]{inputenc}
\usepackage[normalem]{ulem}
\usepackage{tikz}
\usetikzlibrary{shapes,arrows.meta,positioning,matrix}
\tikzset{>=Stealth}
\newcommand\be{\begin{equation}}
\newcommand\ba{\begin{eqnarray}}
\newcommand\ee{\end{equation}}
\newcommand\ea{\end{eqnarray}}
\newcommand\bw{\begin{widetext}}
\newcommand\ew{\end{widetext}}

\newcommand{\order}[1]{\mathcal{O}\left( #1 \right)}

\newcommand{\diff}[1]{\frac{d \, #1}{dr}}

\makeatletter
\newcommand\footnoteref[1]{\protected@xdef\@thefnmark{\ref{#1}}\@footnotemark}
\makeatother

\definecolor{KellyGreen}{RGB}{76,187,23}

\begin{document}
\title{Tidal Deformabilities of Neutron Stars in scalar-Gauss-Bonnet Gravity \\ and Their Applications to Multimessenger Tests of Gravity}
\author{Alexander Saffer}
\affiliation{Department of Physics, University of Virginia, Charlottesville, Virginia 22904, USA}
\author{Kent Yagi}
\affiliation{Department of Physics, University of Virginia, Charlottesville, Virginia 22904, USA}

\date{\today}
\begin{abstract} 

The spacetime surrounding compact objects such as neutron stars and black holes provides an excellent place to study gravity in the strong, non-linear, dynamical regime. 
Here, the effects of strong curvature can leave their imprint on observables which we may use to study gravity. 
Recently, NICER provided a mass and radius measurement of an isolated neutron star 
using x-rays, while LIGO/Virgo measured the tidal deformability of neutron stars through gravitational waves. These measurements can be used to test the relation between the tidal deformability and compactness of neutron stars that are known to be universal in general relativity. Here, we take (shift-symmetric) scalar-Gauss-Bonnet gravity (motivated by a low-energy effective theory of a string theory) as an example and study whether one can apply the NICER and LIGO/Virgo measurements to the universal relation to test the theory. To do so, this paper is mostly devoted on theoretically constructing tidally-deformed neutron star solutions in this theory perturbatively and calculate the tidal deformability for the first time.
We find that the relation between the tidal deformability and compactness remains to be mostly universal for a fixed dimensionless coupling constant of the theory though the relation is different from the one in general relativity.
We also present a universal relation between the tidal deformability \textbf{of} one neutron star and the compactness for another neutron star that can be directly applied to observations by LIGO/Virgo and NICER. For the equations of state considered in this paper, it is still inconclusive whether one can place a meaningful bounds on scalar Gauss-Bonnet gravity with the new universal relations. However, we found a new bound from the mass measurement of the pulsar J0740+6620 that is comparable to other existing bounds from black hole observations.
\end{abstract}
\maketitle

\section{Introduction}
\label{sec:Introduction}
Neutron stars (NSs) represent some of the densest objects in the universe, second only to black holes.  The densities of these stars can reach several times nuclear saturation density ($\rho = 2.8 \times 10^{14} \, {\rm g/cm^3}$) which is greater than any density measurable in a laboratory~\cite{Lattimer:2006xb}.  This immense density coincides with masses on the order of $1.5 \, M_\odot$ and radii around 12--15 km~\cite{Lattimer:2004pg}.  The exact properties of a NS can be found given a specific equation of state (EoS) which determines the relationship between the pressure and density within the star.  Observations of NSs' mass and radius could allow scientists to understand the inner workings of these EoSs better and allow for a more fundamental understanding of nuclear physics.

One way scientists have sought to explore the inner workings of NSs has been with the Neutron star Interior Composition ExploreR (NICER).  This instrument aims to provide inferences of the mass and radius of NSs (whose relation depends sensitively on the underlying EoS) to accuracies never before achieved with other optical telescopes~\cite{10.1117/12.926396,10.1117/12.2056811}, through the use of monitoring the x-ray hotspots on  a rotating NS.  The results obtained via PSR J0030+0451~\cite{Riley:2019yda,Miller:2019cac} placed stringent bounds on the valid EoS~\cite{Raaijmakers:2019qny,Annala:2021gom,Pang:2021jta,Legred:2021hdx}.

Another way of probing the internal structure of NSs is to use gravitational-wave observations, as done by the LIGO/Virgo Collaboration (LVC). Through the binary NS merger event GW170817, LVC measured
the (dimensionless) tidal deformability parameter  ($\Lambda$), which measures the susceptibility of a NS to be deformed by an external tidal field~\cite{Hinderer:2007mb,Chatziioannou:2020pqz}.  The larger $\Lambda$ is, the easier a star will deform.
Observations of the event GW170817 have placed limits on the tidal parameter for a $1.4\,M_\odot$ NS to be $\Lambda_{1.4} = 190^{+390}_{-120}$~\cite{Abbott:2018exr}. 
Once again, this observation has led to stringent bounds on the EoS (see e.g.~\cite{Abbott:2018exr,Annala:2017llu,Raithel:2018ncd,Lim:2018bkq,Bauswein:2017vtn,De:2018uhw,Most:2018hfd,Annala:2019puf,Malik2018,Zack:nuclearConstraints,Carson:2019xxz,Raithel:2019ejc,Chatziioannou:2020pqz}), including joint bounds between x-ray and gravitational waves~\cite{Raaijmakers:2019dks,Zimmerman:2020eho,Jiang:2019rcw,Dietrich:2020efo}.

Due to their large compactness and strong gravitational field strength, NSs are ideal sources to probe not only nuclear physics but also gravity~\cite{stairs,Abbott:2018lct}.
Thus far, all measurements of NS properties have agreed with those predicted by general relativity (GR) for a number of EoSs.  However, GR itself has problems elsewhere.  It has been shown that given a field-theoretical approach, GR is non-renormalizable which may lead to problems in the ultra-violet regime.  The predictions of quantum gravity theories in the low energy limit often show that GR should be modified by additional fields and higher-order curvature scalar terms~\cite{Yagi_2016}.  This hints at the possibility that GR as currently understood is incomplete and may be modified at specific energy levels not yet studied in detail, such as those in the strong gravity regime.

One alternative theory of gravity which has drawn interest from physicists is Einstein-dilaton-Gauss-Bonnet (EdGB) gravity, in which a scalar field (dilaton) is coupled exponentially to the metric.  This theory shows up in the low-energy limit of heterotic string theory~\cite{Berti:2015itd,Zhang:2017unx}, and has been shown to agree with GR in the weak field regime~\cite{Sotiriou:2006pq}.  With this in mind, the next step would then be to examine EdGB in the strong field regime.  Work has already been completed on studying this theory for both BHs~\cite{Kanti:1995vq,Torii:1996yi,Kleihaus:2011tg,Yunes:2011we,Sotiriou:2014pfa,Ayzenberg:2014aka} and NSs~\cite{Pani:2011xm,Kleihaus:2014lba,Doneva:2017duq,Saffer:2019hqn,Charmousis:2021npl}.  These studies have managed to place some limits on the theory, but improvements are necessary as new ways of probing the strong field are undertaken. In this paper, we consider scalar-Gauss-Bonnet (sGB) gravity that generalizes the form of the coupling of the scalar field to the metric and includes EdGB gravity as an example.  As a simplification to this theory, we make the assumption that our deviation from GR is small and we may decouple the theory by linearly coupling our scalar field to the Gauss-Bonnet parameter~\cite{Yagi:2015oca}.

One difficulty in using NSs to test gravity is the degeneracy between uncertainties in nuclear physics and gravitational physics. For example, one can in principle use the relation between the NS mass and radius to probe gravity, though one needs to know the correct EoS beforehand. One way to overcome this is to use certain universal relations that are insensitive to the underlying EoSs~\cite{Yagi365,Yagi:2013awa,Yagi:2016bkt,Doneva:2017jop}. For example, universal relations exist between the NS moment of inertia ($I$), the tidal Love number (Love), and the spin-induced quadrupole moment $Q$~\cite{Yagi365,Yagi:2013awa}. The I-Q relation has been studied in EdGB gravity in~\cite{Kleihaus:2014lba,Kleihaus:2016dui}.

In this paper, we aim to probe sGB gravity by applying universal relations to the NS measurement by NICER and LVC. A similar analysis has been carried out in~\cite{Silva:2020acr} to probe dynamical Chern-Simons gravity~\cite{Alexander:2009tp}. The authors in~\cite{Silva:2020acr} converted the NICER's measurement of the NS compactness to the moment of inertia by using the universal relation between these two quantities assuming GR is the correct theory of gravity. The authors then used the tidal deformability measurement by LVC and the I-Love universal relations to probe the theory. In this paper, we focus on a different universal relation, namely the one between the tidal deformability and compactness (Love-C relation)~\cite{Maselli:2013mva}. Since these quantities are the ones that are directly measured through x-ray and gravitational waves, we do not need to apply an additional universal relation to convert the measurement of one quantity from another.

In order to develop these relationships in sGB gravity, a significant portion of this paper is devoted to computing the NS tidal deformability by constructing tidally-deformed NS solutions in sGB by following a similar procedure to one typically performed in GR.

The tidal deformability is defined as the ratio between the quadrupolar deformation of a body ($Q_{ij}$) and an exterior quadrupolar tidal field ($\mathcal{E}_{ij}$)~\cite{Hinderer:2007mb,Thorne:1997kt}.  This may be computed via an asymptotic expansion of the external metric about a distance much larger than the radius of the star, which allows one to denote the $Q_{ij}$ and $\mathcal{E}_{ij}$ as relating to the coefficients to the $\order{1/r^3}$ and $\order{r^2}$ terms respectively.  For sGB, we compute these quantities assuming that the tidal field and the sGB corrections are small, and solving the field equations perturbatively in terms of both a tidal parameter $\epsilon$ and the sGB parameter $\alpha$ that characterizes the coupling between the scalar field and the metric. Once we have tidal deformability calculations in hand, we compare these theoretical predictions with various NS observations to constrain sGB gravity.

\begin{figure}
    \includegraphics[width=8.5cm]{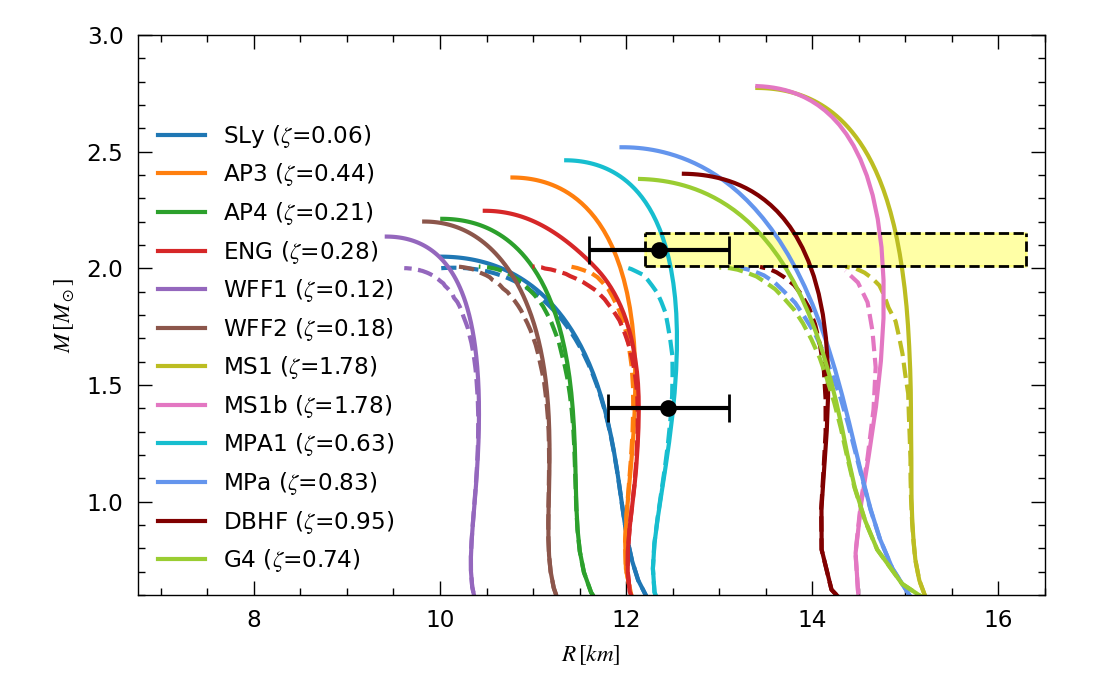}
    \caption{The mass-radius curves in GR (solid) and in sGB (dashed) with the maximum value of $\zeta$ for each EoS allowed to support a $2.01M_\odot$ NS, which corresponds to lower mass bound for J0740+6620 in~\cite{Cromartie:2019kug,Fonseca:2021wxt,Riley:2021pdl,Miller:2021qha} for a selection of EoS.
    We terminate the curves at the maximum mass of the NS for the corresponding $\zeta$.
    The yellow box is representative of mass and radius bounds inferred for J0740+6620 with 1-$\sigma$ errors~\cite{Miller:2021qha}. The two black dots with error bars correspond to 1-$\sigma$ bounds placed on the radius of NSs with masses $1.4 M_\odot$ and $2.08 M_{\odot}$ from a recent NICER analysis by combining measurements of neutron stars through x-ray, radio and gravitational waves~\cite{Miller:2021qha}.
}
    \label{fig:MRcurve_max_mass}
\end{figure}

We find the following main results. Figure~\ref{fig:MRcurve_max_mass} presents the mass-radius relation (for isolated, non-tidally-deformed NSs) in GR and in sGB. Here, $\zeta$ is a new coupling constant in sGB gravity where we make $\alpha$ dimensionless by the mass and radius of a neutron star (we define this quantity properly in Eq.~\eqref{eq:NewZetaDef}).  Notice that the maximum mass decreases in sGB gravity as was first found in~\cite{Saffer:2019hqn}.  In the figure, we have chosen the dimensionless coupling constant that can marginally support a 2.01$M_\odot$ NS, which is the lower mass bound for J0740+6620~\cite{Cromartie:2019kug,Fonseca:2021wxt}. 
We also assume the correction to the Shapiro delay used to infer the mass in~\cite{Cromartie:2019kug,Fonseca:2021wxt} is not corrected by the sGB corrections.
This assumption is based on the results of previous work~\cite{Saffer:2019hqn}, where it was found that perturbations to the metric of a NS occur at $\order{M^7/r^7}$ where $M$ is the stellar mass while $r$ is the distance from the star.  Thus, we expect the influence on the Shapiro delay from sGB corrections to be highly suppressed.
Based on this, we can place bounds on sGB gravity that is EoS dependent. Choosing the stiffest EoS (MS1 and MS1b), we found a bound on the dimensionful coupling constant $\sqrt{\alpha} < 1.29$km, which provides the most conservative bound out of the EoS considered in this paper. This new bound  is comparable to other existing bounds from BH observations summarized in Table~\ref{tab:summary}, and in fact is the strongest if we do not account for the bound from GW190814~\cite{Wang:2021yll} whose secondary object is uncertain.

\begin{figure}
 	\includegraphics[width=\linewidth]{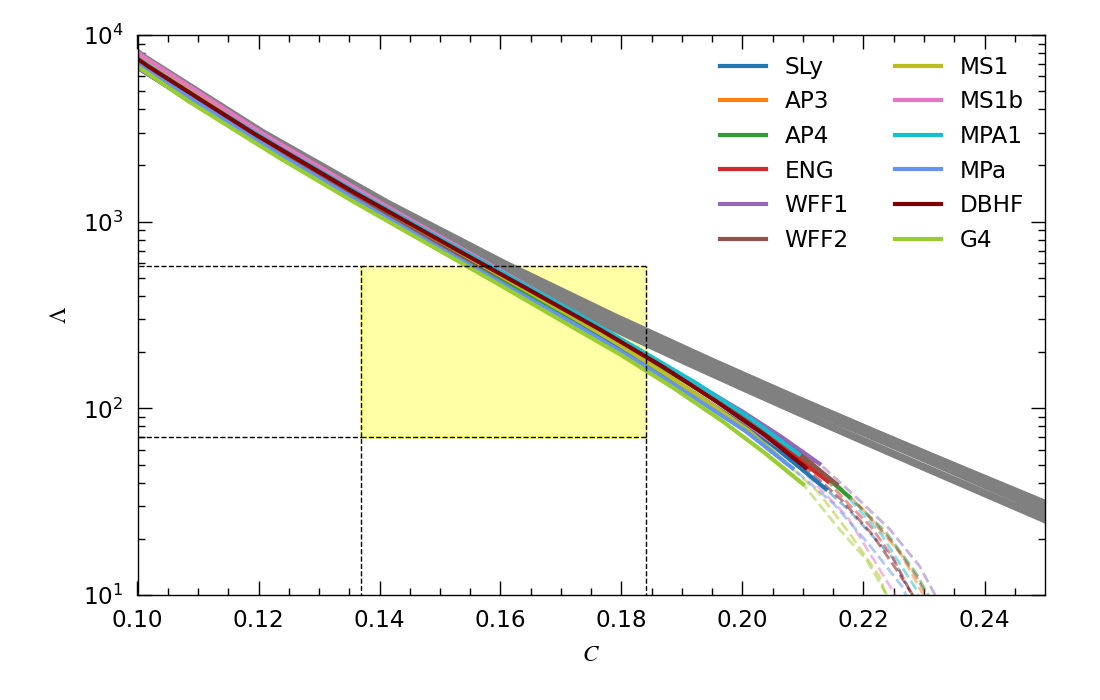}
     \caption{
Relation between the tidal deformability and compactness of neutron stars with various EoSs.  GR solutions are presented as solid grey lines, while sGB solutions with the dimensionless coupling constant of $\zeta=0.5$ are given as colored lines.
The dashed lines correspond to the continuation of the sGB solution following $\Lambda_{\rm sGB} < 0.5 \Lambda_{\rm GR}$, where we take the small coupling approximation to no longer be valid.
The yellow box corresponds to bounds placed on a $1.4 M_\odot$ NS with the $\Lambda$ measurement from GW170817 by LIGO/Virgo ($\Lambda_{1.4} = 190^{+390}_{-120}$)~\cite{Abbott:2018exr} and the compactness one from NICER ($C_{1.4} = 0.159^{+0.025}_{-0.022}$)~\cite{Silva:2020acr}. }
     \label{fig:lambdaCcurve}
\end{figure}

Regarding tidally-deformed NSs, we first derive sGB correction to the dimensionless tidal deformability $\Lambda$ (Eq.~\eqref{eq:CorrectedLove}). The correction arises from two parts, one on the dimensionful tidal deformability and another on the stellar mass used to normalize the tidal deformability. These corrections both enter at quadratic order in the sGB coupling constant $\alpha$. We next find that in general, the universal relations between $\Lambda$ and $C$ remain relatively EoS insensitive, though the EoS variation increases slightly from the GR case (see Fig.~\ref{fig:lambdaCcurve}). The deviation from GR in the universal relation increases for NSs with larger compactnesses, as the stellar curvature gets larger and the sGB correction becomes larger. Unfortunately, this deviation is small and not detectable by current measurement standards for NSs with $\sim 1.4M_\odot$ from GW170817 and J0030+0451. 

So far, studies on testing GR through multimessenger observations via universal relations~\cite{Yagi365,Yagi:2013awa,Silva:2020acr} have focused on combining  measurements of two different quantities for the same NS mass (e.g. $1.4M_
\odot$) as the universal relations were constructed as a sequence of a single NS. To go beyond this, we study the relation between $\Lambda$ and $C$ for NSs with different masses. To be more specific, we compared the $\Lambda$ of a $1.4 M_\odot$ NS to the $C$ of a $2.08 M_\odot$ NS. 
We find that such a relation is EoS universal to a fractional variation of $\sim 10\%$ and while this test seems useful, this examination failed to provide any meaningful bounds on the theory in question.  We believe that more EoS may be needed to further investigate this approach.

\renewcommand{\arraystretch}{1.2}
\begin{table*}[tb]
\begin{centering}
\begin{tabular}{r|c|c|c|c|c}
\hline
\hline
\noalign{\smallskip}
\multirow{2}{*}{}  & \multirow{2}{*}{LMXB} &
 \multicolumn{2}{c|}{BBH} &  \multicolumn{2}{c}{\textbf{NS (this work)}} \\
\cline{3-6}
&  & O1--O2 & O1--O3 & max mass & $\Lambda$--$C$   \\
\hline
$\sqrt{\alpha}$ [km]& 1.9~\cite{Yagi:2012gp} & 5.6~\cite{Nair:2019iur}, 1.85~\cite{Yamada:2019zrb}, 4.3~\cite{Tahura:2019dgr} &  1.7~\cite{Perkins:2021mhb}, 4.5~\cite{Wang:2021yll},  (0.4)~\cite{Wang:2021yll}  &\textbf{1.29} & \textbf{---} \\
\noalign{\smallskip}
\hline
\hline
\end{tabular}
\end{centering}
\caption{Astrophysical bounds on (linearly-coupled) sGB gravity. We show bounds from a LMXB, binary black holes (BBHs) and NSs. For BBH, the bounds come from gravitational wave observations and we show bounds using events during O1--O2 runs and O1--O3 runs separately. The one in brackets come from GW190814 assuming that it is a BBH, which has some uncertainty and the bound becomes much weaker if it is a NS-BH binary. For NS, we present the bound from the NS maximum mass. The one from the universal relation between the tidal deformability and compactness ($\Lambda$--$C$) is still inconclusive and needs a further study.
}
\label{tab:summary}
\end{table*}

The structure of the paper is as follows: In Sec.~\ref{sec:sGB} we explain the details of sGB theory.  Section~\ref{sec:EdGBLoveFullSection} presents our derivation of the sGB correction to the tidal deformability parameter.  Section~\ref{sec:Astro} will discuss the implications of these calculations for various EoSs and how they relate to the results from NICER and LIGO.
In Sec.~\ref{sec:Discussion} we discuss the results of our analysis and compare what we find with observations.
Throughout this paper, we will make use of the metric signature $(-,+,+,+)$ as presented in~\cite{Misner:1974qy} and units $c=G=1$.

\section{scalar-Gauss-Bonnet Gravity}
\label{sec:sGB}
In this section, we will detail the decomposition of sGB from EdGB, the action of the theory, and the field equations we will use throughout this paper.

\subsection{Action}
\label{subsec:sGBAction}
We begin with the basics of sGB gravity as explained in~\cite{Yagi_2016}.  The action for sGB is
\begin{equation}
    \label{eq:EdGBAction}
    S=\int \sqrt{-g} \left[ \kappa R - \frac{1}{2} \nabla_\mu \varphi \nabla^\mu \varphi + \alpha f(\varphi) \mathcal{G} + 2U(\varphi)\right] d^4x\,,
\end{equation}
where $\kappa=\left(16 \pi\right)^{-1}$, $\varphi$ is a canonical scalar field with potential $U(\varphi)$, $f(\varphi)$ is a functional coupling of the scalar field to the metric with coupling strength $\alpha$ (that has a unit of length squared in the $c=G=1$ unit and when $\varphi$ is dimensionless), and $\mathcal{G}$ is the Gauss-Bonnet constant defined as
\begin{equation}
    \label{eq:GaussBonnetTerm}
    \mathcal{G} \equiv R^2 - 4R_{\mu \nu}R^{\mu \nu} + R_{\mu \sigma \nu \tau}R^{\mu \sigma \nu \tau}\,.
\end{equation}
There are a number of forms the functional, $f(\varphi)$, can take (see~\cite{Antoniou:2017hxj,Doneva:2017bvd,Antoniou_2018,Doneva_2018,Silva:2017uqg,Silva:2018qhn}). For example, $f(\varphi) = e^{-\gamma \varphi}$ for a constant $\gamma$ corresponds to Einstein-dilaton Gauss-Bonnet gravity that arises as a low-energy limit of a string theory~\cite{Maeda:2009uy}.

In this paper, we choose to work in the so called decoupling limit of the theory~\cite{Yagi_2016} with a massless dilaton such that $U(\varphi)=0$.  From this, our functional is Taylor expanded about an asymptotic value of $\varphi$ at infinity that we take to be 0\footnote{As we explain later, a linearly-coupled sGB has shift-symmetry and thus the asymptotic value of the scalar field is irrelevant.} in the form
\begin{equation}
    \label{eq:DecoupleLimit}
    f(\varphi) = f(0) + f_{,\varphi}(0) \varphi + \order{\varphi^2}\,.
\end{equation}
The first term in Eq.~\eqref{eq:DecoupleLimit} contributes no information to the dynamics of the problem.  This is due to the fact that $\mathcal{G}$ is a topological constant and thus when the first term is considered, it yields a boundary term in the action which does not contribute to the dynamics of the problem.  The second term in Eq.~\eqref{eq:DecoupleLimit} will be used to convert our full action to \textit{decoupled dynamical Gauss-Bonnet gravity} or shift-symmetric sGB gravity (which we refer simply to sGB gravity in this paper).  
The final action of sGB gravity that we consider in this paper is
\begin{equation}
    \label{eq:sGBAction}
    S=\int \sqrt{-g} \left[ \kappa R - \frac{1}{2} \nabla_\mu \varphi \nabla^\mu \varphi + \alpha \varphi \mathcal{G} \right] d^4x\,.
\end{equation}
Note that the theory becomes invariant under the transformation $\varphi \rightarrow \varphi + c$~\cite{Sotiriou:2013qea,Sotiriou:2014pfa,Saravani:2019xwx}, where $c$ is a constant.  This is an example of shift-symmetric Horndeski gravity~\cite{Kobayashi:2011nu}. 
In Eq.~\eqref{eq:sGBAction}, we have absorbed the constant $f_{,\varphi}(0)$ into the coupling parameter $\alpha$. 

The coupling parameter has been constrained to be $\sqrt{\alpha} \lesssim \mathcal{O}(1)$km from black hole observations through x-rays~\cite{Yagi:2012gp} and gravitational waves~\cite{Nair:2019iur,Yamada:2019zrb,Tahura:2019dgr,Wang:2021yll,Perkins:2021mhb} (see Table~\ref{tab:summary}). Both low-mass x-ray binary (LMXB) and gravitational wave sources were used to probe the existence of scalar dipole radiation, that is present in sGB gravity, through the measurement of the orbital decay rate. Thus, one can probe sGB gravity with dynamical spacetime through these systems (as opposed to static spacetimes from the maximum mass of neutron stars found in this paper) and the length scale being probed is roughly the size of BHs (though the bounds depend also on how accurately one can measure the orbital decay rate).

Given that we are neglecting curvature interactions higher than cubic order in the action, we treat the theory in Eq.~\eqref{eq:sGBAction} as an effective theory and work in the small coupling approximation. Namely, we assume $\bar{\zeta} \ll 1$, where $\bar{\zeta}$ is the dimensionless coupling constant given by~\cite{Yagi:2013mbt,Gupta:2017vsl,Saffer:2019hqn}
\begin{equation}
\label{eq:bar_zeta}
    \bar{\zeta} \equiv \frac{16\,\pi \alpha^2}{L^4} = \frac{16\,\pi \alpha^2 M_0^2}{R_0^6}\,,
\end{equation}
where $L \equiv \sqrt{R_0^3/M_0}$ characterizes the curvature length of a NS with $M_0$ and $R_0$ representing the stellar mass and radius in GR.  For the duration of this work, we scale this quantity by the compactness such that
\begin{equation}
\label{eq:NewZetaDef}
\zeta \equiv \frac{\bar{\zeta}}{C_0^6}  = \frac{16\,\pi \alpha^2}{M_0^4}\,,
\end{equation}
which is another dimensionless coupling parameter used widely in the literature.
We aim to construct tidally-deformed NS solutions in linearly-coupled sGB gravity under the small coupling approximation valid to $\mathcal{O}(\zeta)$ or $\mathcal{O}(\alpha^2)$.

\subsection{Field Equations}
\label{subsec:FieldEquations}
The field equations may be found by varying the action with respect to our dynamical fields, the metric $g^{\mu \nu}$ and $\varphi$.  This leads to
\begin{subequations}
\begin{align}
\label{eq:metric_FE}
G_{\mu \nu}  &= - \frac{\alpha}{\kappa}\mathcal{K}_{\mu \nu} + \frac{1}{2\kappa}\left(T_{\mu \nu}^{\rm m} + T_{\mu \nu}^\varphi\right)\,, \\
\label{eq:field_FE}
\Box \varphi &= -\alpha \, \mathcal{G}\,,
\end{align}
\end{subequations}
where $G_{\mu \nu}$ is the typical Einstein tensor, $\mathcal{K}_{\mu \nu}$ is given by
\begin{align}
\mathcal{K}_{\mu \nu} &= -2R \nabla_\mu \nabla_\nu \varphi +2\left(g_{\mu \nu}R-2R_{\mu \nu}\right) \Box \varphi  \nonumber \\
&\quad + 8R_{\gamma(\mu}\nabla^\gamma \nabla_{\nu)}\varphi - 4g_{\mu \nu}R^{\gamma \delta}\nabla_\gamma \nabla_\delta\varphi \nonumber \\
&\quad + 4R_{\mu \gamma \nu \delta}\nabla^\gamma \nabla^\delta \varphi\,,
\end{align}
the energy-momentum tensor for the scalar field is
\begin{equation}
T^\varphi_{\mu \nu} = \nabla_\mu \varphi \nabla_\nu \varphi - \frac{1}{2}g_{\mu \nu} \nabla_\gamma \varphi \nabla^\gamma \varphi \,,
\end{equation}
and we define $\Box \equiv \nabla_\mu \nabla^\mu$.  We will assume that the matter we are dealing with inside of the NS can be described as a perfect fluid, and so define the matter stress tensor in Eq.~\eqref{eq:metric_FE} as
\begin{equation}
\label{eq:MatterSET}
T^{\mu \nu}_{\rm m} = \left(\rho + p \right)u^\mu u^\nu + p\,g^{\mu \nu}\,,
\end{equation}
with pressure $p$, energy density $\rho$, and four-velocity of the fluid $u^\mu$.  
This four-velocity is also forced to obey the contraint that this fluid is timelike such that $u_\mu u^\mu = -1$.  Equation~\eqref{eq:MatterSET} satisfies the conservation $\nabla^\mu T_{\mu \nu}^{\rm m}=0$.

\section{Constructing Tidally-deformed Neutron Stars}
\label{sec:EdGBLoveFullSection}

In this section, we explain in detail how to construct tidally-deformed neutron star solutions. We begin by reviewing the ansatz for the metric, matter and scalar field. We next derive tidal perturbation equations in GR, present the asymptotic behavior of the solutions and extend this formulation to sGB gravity. We finally describe how to compute the tidal Love number and tidal deformability for neutron stars.

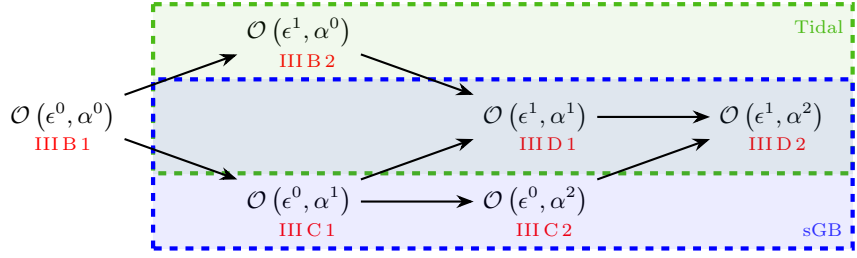
\begin{figure*}
    \centering
    \begin{tikzpicture}[
	thick,
	box/.style={
		draw, 
		text width=6em, 
		align=center,
		minimum height=1cm
	}
	]
    \begin{scope}[node distance=0.5cm and 1.5cm]
 		\node (GR) {$\order{\epsilon^0,\alpha^0}$};
 		\node (o10) [above right=of GR] {$\order{\epsilon^1,\alpha^0}$};
 		\node (o11) [below right=of o10] {$\order{\epsilon^1,\alpha^1}$};
		\node (o01) [below right=of GR] {$\order{\epsilon^0,\alpha^1}$};
 		\node (o02) [right=of o01] {$\order{\epsilon^0,\alpha^2}$};
 		\node (o12) [right=of o11] {$\order{\epsilon^1,\alpha^2}$};
 		
 		\node[color=black,opacity=1] (e.text) at (0cm,-0.35cm) {\scriptsize \ref{sec:background}};
 		
		\node[color=black,opacity=1] (e.text) at (3.25cm,0.75cm) {\scriptsize \ref{subsubsec:GRTidal}};
		\node[color=black,opacity=1] (e.text) at (3.25cm,-1.5cm) {\scriptsize \ref{subsubsec:eps0_alpha1}};
		
		\node[color=black,opacity=1] (e.text) at (6.45cm,-0.35cm) {\scriptsize \ref{subsubsec:ScalarLoveCorrection}};
		\node[color=black,opacity=1] (e.text) at (6.4cm,-1.5cm) {\scriptsize \ref{subsubsec:eps0_alpha2}};
		
		\node[color=black,opacity=1] (e.text) at (9.5cm,-0.35cm) {\scriptsize \ref{subsubsec:MetricLoveCorrection}};
		
		\draw[KellyGreen, ultra thick, dashed] (1.2cm,1.5cm) rectangle (10.5cm,-0.75cm);
		\fill [KellyGreen,opacity=0.075] (1.2cm,1.5cm) rectangle (10.5cm,-0.75cm);
		\node[color=KellyGreen,opacity=1] (e.text) at (10.05cm,1.2cm) {\scriptsize Tidal};
					
		\draw[blue, ultra thick, dashed] (1.2cm,-1.75cm) rectangle (10.5cm,0.5cm);
		\fill [blue,opacity=0.075] (1.2cm,-1.75cm) rectangle (10.5cm,0.5cm);
		\node[color=blue,opacity=0.8] (e.text) at (10.1cm,-1.5cm) {\scriptsize sGB};
	\end{scope}
	
	\path [->]
	(GR) edge (o10)
	(GR) edge (o01)
	(o10) edge (o11)
	(o11) edge (o12)
	(o01) edge (o02)
	(o02) edge (o12)
	(o01) edge (o11)
	;
\end{tikzpicture}
    \caption{Flowchart illustrating the method order used in this paper.  Relevant sections for specific $\order{\epsilon^m,\alpha^n}$ solutions are included.  The colored boxes indicate which aspects of the calculations are being considered as deviations from the unperturbed GR solution: tidal perturbations are shown in green, while sGB corrections are shown in blue.  The overlap of the areas represents the new work in this paper.}
    \label{fig:FlowChart}
\end{figure*}

\subsection{Metric, Matter and Scalar Field Decomposition}
\label{subsec:MetricDecomp}

We begin by explaining the metric ansatz for tidal perturbation in sGB gravity. First, the background, static, spherically symmetric line element is given by
\begin{equation}
    \label{eq:LineElement1}
    ds^2 = -e^{\tau(r)} dt^2 + e^{\sigma(r)} dr^2 + r^2 d\Omega^2\,,
\end{equation}
where $d\Omega^2 = d\theta^2 + \sin^2 \theta \, d\phi^2$.  We next choose to decompose our metric terms as a series involving order-by-order perturbations.  
There are two perturbations to consider.  For the first, denoted as $\epsilon$, we use an order keeping parameter to represent the tidal deformability.  For our second perturbation, we make use of $\alpha$ to represent our sGB perturbation paramater as shown in Eq.~\eqref{eq:sGBAction}.
We may therefore write our metric as a perturbation series to $\order{\epsilon,\alpha^2}$.\footnote{Reference~\cite{Mignemi_1993} showed that the leading order correction to the metric in sGB is of $\order{\alpha^2}$.}
Assuming small deviations from GR (i.e. $\epsilon \ll 1$ and $\alpha \ll L^2$), we may expand the metric given in Eq.~\eqref{eq:LineElement1} using 
$\tau(r) = \tau_{00}(r) + \epsilon \tau_{10}(r) \,  Y_{2m}(\theta,\phi) + \alpha^2 \tau_{02}(r) + \epsilon \alpha^2 \tau_{12}(r) Y_{2m}(\theta,\phi) $ (with similar forms for the other metric functions). Here $Y_{lm}$ are spherical harmonics and we fix the angular dependence to the $l=2$ spherical harmonics since we are interested in quadrupolar tidal deformations.  For example, the $(t,t)$ component of the metric is given by 
\begin{align}
g_{tt} = - e^\tau =& - e^{\tau_{00} + \epsilon \tau_{10}Y_{2m} + \alpha^2 \tau_{02} + \epsilon \alpha^2 \tau_{12}Y_{2m}} \nonumber \\
=& - e^{\tau_{00}} \,e^{\epsilon \tau_{10}Y_{2m}}  \,e^{\alpha^2 \tau_{02}} \, e^{\epsilon \alpha^2 \tau_{12}Y_{2m}} \nonumber \\
=& - e^{\tau_{00}} \left[1+ \epsilon \, \tau_{10} \,  Y_{2m}(\theta,\phi)\right]\left[1+\alpha^2 \, \tau_{02}\right] \nonumber \\
& \times\left[ 1+ \epsilon \alpha^2 \, \tau_{12} Y_{2m}(\theta,\phi)\right] + \mathcal{O}(\epsilon^2,\alpha^4)\,.
\end{align}
Following this, the metric ansatz is given by
\begin{widetext}
\begin{align}
    \label{eq:ExpandedMetric}
    ds^2 &= -e^{\tau_{00}(r)}\left[1+ \epsilon \, \tau_{10}(r) \,  Y_{2m}(\theta,\phi)\right]\left[1+\alpha^2 \, \tau_{02}(r)\right] \left[ 1+ \epsilon \alpha^2 \, \tau_{12}(r) Y_{2m}(\theta,\phi)\right] dt^2 \nonumber \\
    & \quad + e^{\sigma_{00}(r)}\left[1+ \epsilon \, \sigma_{10}(r) \,  Y_{2m}(\theta,\phi)\right]\left[1+\alpha^2 \, \sigma_{02}(r)\right] \left[ 1+ \epsilon \alpha^2 \, \sigma_{12} (r)Y_{2m}(\theta,\phi)\right] dr^2 \nonumber \\
    & \quad + r^2 \left(1+ \epsilon \, K_{10} \,  Y_{2m} + \epsilon \alpha^2 \, K_{12} Y_{2m}\right) d\Omega^2 +\mathcal{O}(\epsilon^2,\alpha^3) \nonumber \\
    &= -e^{\tau_{00}}\left[1+ \epsilon \, \tau_{10} \,  Y_{2m} +\alpha^2 \, \tau_{02}+ \epsilon \alpha^2 \, (\tau_{12} + \tau_{10} \tau_{02}) Y_{2m}\right] dt^2 \nonumber \\
    & \quad + e^{\sigma_{00}}\left[1+ \epsilon \, \sigma_{10} \,  Y_{2m}+\alpha^2 \, \sigma_{02}+ \epsilon \alpha^2 \, (\sigma_{12} + \sigma_{10} \sigma_{02}) Y_{2m}\right] dr^2 \nonumber \\
    & \quad + r^2 \left(1+ \epsilon \, K_{10} \,  Y_{2m} + \epsilon \alpha^2 \, K_{12} Y_{2m}\right) d\Omega^2 +\mathcal{O}(\epsilon^2,\alpha^3) \,.
\end{align}
\end{widetext}
Notice that there is no $K_{02}$ as it can be gauged away~\cite{Saffer:2019hqn}.
 We will also only consider the axi-symmetric ($m=0$) modes, which allow us to write the explicit form of the angular dependence as\footnote{This choice is just for simplicity as the tidal deformability is known to be independent of $m$~\cite{Hinderer:2007mb}.}
\begin{equation}
    Y_{20} = \frac{1}{4}\sqrt{\frac{5}{\pi}} \left[ 3 \cos^2\left( \theta \right) - 1 \right]\,.
\end{equation}In addition to the background functions $\tau_{ab}(r)$ and $\sigma_{ab}(r)$ at different orders, we introduce a new metric term $K_{ab}(r)$ to include radial dependence in the angular component of the tidal perturbations. The first index on radial functions counts the order of tidal perturbation $\epsilon$ while the second index counts the order of $\alpha$.  

We will also expand our scalar field and matter components in terms of $\epsilon$ and $\alpha$ as well\footnote{The $\order{\alpha}$ components of the pressure and density do not contribute to the field equations~\cite{Saffer:2019hqn}. }
\begin{subequations}
\begin{align}
\label{eq:varphiExpand}
    \varphi &= \varphi_{00} + \alpha \, \varphi_{01} + \epsilon \alpha \, \varphi_{11} +  \order{\epsilon,\alpha^2}\,, \\
    \label{eq:pressExp}
    p &= p_{00} + \epsilon \, p_{10} + \alpha^2 \, p_{02}  +  \epsilon \alpha^2 p_{12} + \order{\epsilon,\alpha^3}\,, \\
    \label{eq:rhoExp}
     \rho &= \rho_{00} + \epsilon \, \rho_{10} + \alpha^2 \, \rho_{02} + \epsilon \alpha^2 \rho_{12} +\order{\epsilon,\alpha^3}\,.
\end{align}
\end{subequations}
The purpose of this expansion will be clear as we progress through the derivation.

\subsection{General Relativity}
\label{subsec:GR}

We now derive tidal perturbation equations in GR.
For our analysis of GR, we will restrict ourselves to the metric under the assumption $\alpha=0$:
\begin{align}
    \label{eq:ExpandedMetricGR}
    ds^2_{\rm GR} &= -e^{\tau_{00}}\left(1+ \epsilon \, \tau_{10} \,  Y_{2m}\right) dt^2 \nonumber \\
    & \quad + e^{\sigma_{00}}\left(1+ \epsilon \, \sigma_{10} \,  Y_{2m}\right)dr^2 \nonumber \\
    & \quad + r^2 \left(1+ \epsilon \, K_{10} \,  Y_{2m}\right) d\Omega^2\,.
\end{align}
Equation~\eqref{eq:ExpandedMetricGR} is the same metric as in~\cite{Hinderer:2007mb} (see also~\cite{Damour:2009vw,Binnington:2009bb} for relativistic formulation of tidal perturbations in GR), so we will take the same approach to resolving the metric components.  

\subsubsection{Background at $\mathcal{O}(\epsilon^0,\alpha^0)$}
\label{sec:background}

At $\order{\epsilon^0}$, we recover the standard Tolman–Oppenheimer–Volkoff (TOV) equations.  The $(t,t)$ and $(r,r)$ components of Eq.~\eqref{eq:metric_FE} yield 
\begin{subequations}
\begin{align}
\label{eq:m0_eq}
\diff{m_0}&= 4 \pi \rho_{00} r^2\,, \\
\label{eq:tau00_eq}
    \diff{\tau_{00}} &= \frac{2 \left( 4 \pi p_{00} r^3 +m_0 \right)}{r^2 \left(1 - \frac{2m_0}{r}\right)}\,,
\end{align}
\end{subequations}
where we have introduced the mass function 
\begin{equation}
e^{-\sigma_{00}} = \left(1-\frac{2m_0}{r} \right)\,.
\end{equation}
Additionally, conservation of the matter stress tensor $\nabla_\mu T_{\rm m}^{\mu \nu}=0$ gives
\begin{equation}
    \label{eq:pressure_gr}
  \diff{p_{00}} = - \frac{\left(\rho_{00} + p_{00}\right) \left( 4 \pi p_{00} r^3 + m_0 \right)}{r^2 \left( 1 - \frac{2m_0}{r}\right)}\,.
\end{equation}

We numerically construct the interior solution as follows.
At the center of the star, we may Taylor expand our functions $(m_0,\tau_{00}, p_{00})$ about $r=0$ and choose some initial small $r_0$ as our starting point.  Here, we choose some initial density and find the corresponding pressure through the EoS we are using.  This gives us our initial conditions for the differential equations we need to solve.  We choose to terminate the integration of Eqs.~\eqref{eq:m0_eq} and~\eqref{eq:pressure_gr} when the pressure decreases by a factor of $10^{-11}$ from the initial pressure determination; this is the radius we call $R_0$. We next choose a trial central value for $\tau_{00}$ and solve Eq.~\eqref{eq:tau00_eq} to construct a trial solution for $\tau_{00}$.  
The EoSs we use throughout this paper are AP3~\cite{Akmal:1998cf}, AP4~\cite{Akmal:1998cf}, ENG~\cite{Engvik:1995gn}, DBHF~\cite{Katayama:2014gca}, G4~\cite{Lackey:2005tk}, MPa~\cite{Yamamoto:2014jga}, MPA1~\cite{MUTHER1987469}, MS1 and MS1b~\cite{Mueller:1996pm}, SLy~\cite{Akmal:1998cf}, WFF1~\cite{PhysRevC.38.1010}, and WFF2~\cite{PhysRevC.38.1010}.

For the exterior of the star, we may solve the above equations assuming $\rho_{00} = p_{00} = 0$.  We determine the metric components to be 
\begin{subequations}
\begin{align}
    \sigma^{\rm ext}_{00} &= - \ln \left( 1- \frac{2M_0}{r}\right)\,, \\
    \tau^{\rm ext}_{00} &=  \ln \left( 1- \frac{2M_0}{r}\right)\,,
\end{align}
\end{subequations}
where $M_0$ is the total mass of the star enclosed in stellar radius $R_0$, and can be found via boundary matching with the interior solutions of the star; $M_0 = m_0(R_0) $. This also determines the exterior solution for $\tau_{00}$, which can be used to obtain the correct interior solution for $\tau_{00}$ by shifting the trial interior solution by a constant. The latter is determined through the matching of the interior and exterior solutions for $\tau_{00}$ at the surface.

\subsubsection{Tidal Perturbation at $\mathcal{O}(\epsilon^1,\alpha^0)$}
\label{subsubsec:GRTidal}

At $\order{\epsilon}$, we are able to solve for the metric components (or their derivatives) through manipulation of the field equation components.  Subtracting the $(\phi,\phi)$ component of Eq.~\eqref{eq:metric_FE} from the $(\theta,\theta)$ component gives us the relation
\begin{equation}
    \label{eq:LoveMetricRelation1}
    \sigma_{10} = - \tau_{10}\,.
\end{equation}
Additionally, the $(r,\theta)$ component provides the relation
\begin{equation}
    \label{eq:LoveMetricRelation2}
    \diff{K_{10}} = - \left(\diff{\tau_{00}}\right) \tau_{10} - \diff{\tau_{10}}\,.
\end{equation}
Conservation of the matter-stress energy tensor also yields two relations:
\begin{subequations}
\label{eq:SetConserve1}
\begin{align}
    p_{10} &= - \frac{\sqrt{5 \pi}}{4 \pi} \left(p_{00} + \rho_{00} \right) \tau_{10} \,,\\
    \rho_{10} &= \frac{\sqrt{5 \pi}r}{4 \pi} \left( \frac{r-2m_0}{4\pi p_{00}r^3+m} \right) \left(\diff{\rho_{00}}\right) \tau_{10}\,.
\end{align}
\end{subequations}
Equations~\eqref{eq:LoveMetricRelation1}, ~\eqref{eq:LoveMetricRelation2}, and~\eqref{eq:SetConserve1} allow us to rewrite the field equations in terms of only a single metric component $\tau_{10}$.  Considering only axisymmetric solutions, we may take the difference between the $(t,t)$ and $(r,r)$ field equations to obtain~\cite{Hinderer:2007mb}
\begin{widetext}
\begin{align}
\label{eq:HindererEquation}
    & \frac{d^2 \, \tau_{10}}{dr^2} + \left[ \frac{2}{r} + e^{\sigma_{00}} \left(\frac{2m_0}{r^2} + 4 \pi r (p_{00}-\rho_{00})\right)\right] \diff{\tau_{10}} \nonumber \\ 
    & \quad - \left[ \frac{6}{r^2} e^{\sigma_{00}} - 4 \pi e^{\sigma_{00}} \left(5\rho_{00} + 9p_{00} +(\rho_{00} + p_{00})\frac{d \, \rho_{00}}{dp_{00}} \right) + \left( \diff{\tau_{00}}\right)^2\right] \tau_{10}=0\,.
\end{align}
\end{widetext}

The interior solution to Eq.~\eqref{eq:HindererEquation} can be found by forcing regularity at the center of the star, which yields the initial condition
\begin{equation}
\label{eq:tau10InteriorExpansion}
    \tau^{\rm int}_{10}(r) \sim a_0 r^2 + \order{r^4}\,,
\end{equation}
where $a_0$ is an integration constant.
The exterior solution for Eq.~\eqref{eq:HindererEquation} is solved by assuming $\rho_{00}=p_{00}=0$ and solving accordingly.  The resulting solution takes the form 
\begin{align}
\label{eq:LoveExtSolution}
    \tau_{10}^{\rm ext} &= c_1 \left( \frac{r}{M_0}\right)^2 \left(1-\frac{2M_0}{r} \right) \left[\frac{3}{2} \ln \left( \frac{r}{r-2M_0}\right) \right.\nonumber \\
    &\left. \quad - \frac{M_0(M_0-r)(2M_0^2+6M_0r-3r^2)}{r^2(2M_0-r)^2}\right] \nonumber \\
    &\quad + 3c_2 \left( \frac{r}{M_0}\right)^2 \left(1-\frac{2M_0}{r}\right)\,,
\end{align}
where $c_1$ and $c_2$ are integration constants which can be solved for at the boundary of the star in terms of the interior initial condition $a_0$.
Taking the limit of this expression as $r \rightarrow \infty$ gives us the series
\begin{align}
\label{eq:tau10ExteriorExpansion}
    \tau^{\rm ext}_{10}(r \rightarrow \infty) &\approx \frac{8}{5}c_1 \left(\frac{M_0}{r}\right)^3 + \order{\frac{1}{r^4}} \nonumber \\
    &\quad + 3 c_2 \left( \frac{r}{M_0}\right)^2 + \order{r}\,,
\end{align}
which has a direct correlation to the external quadruoplar field ($\order{r^2}$) and the  body's quadrupole moment ($\order{1/r^3}$).  

We will later need $K_{10}$ to find the tidal perturbations in sGB gravity.  This quantity can be found by looking soley at the $(r,r)$ field equation of Eq.~\eqref{eq:metric_FE} at $\order{\epsilon^1,\alpha^0}$.  Making use of the GR solutions presented in Sec.~\ref{sec:background}, as well as Eqs.~\eqref{eq:LoveMetricRelation1}--\eqref{eq:SetConserve1} we may simplify the $(r,r)$ equation to solve for $K_{10}$ and find it to be:
\begin{align}
    \label{eq:K0Solution}
    K_{10} &= \frac{1}{2r \left( r-2m_0\right)} \left\{ 4 r  \left(2 m_0 - r\right) \left(p_{00} \pi r^3 + \frac{m_0}{4}\right) \diff{\tau_{10}} \right. \nonumber \\
    &\left. \quad -32 \left[ -\frac{m_0^2}{16} + \left(\frac{3 \pi}{4} p_{00} r^3 +\frac{\pi}{4} \rho_{00} r^3 - \frac{1}{16} r\right) m_0 \right. \right. \nonumber \\
    & \left. \left. \quad + r^2 \left(\pi^2 p_{00}^2 r^4 - \frac{\pi}{8} p_{00} r^2 - \frac{\pi}{8} r^2 \rho_{00} + \frac{1}{16}\right)\right]\tau_{10}\right\}\,.
\end{align}
The corresponding interior and exterior solutions may be found with the appropriate substitutions.

\subsection{Scalar-Gauss-Bonnet Corrections: Background}
\label{subsec:EdGB}
Let us next extend the GR formulation reviewed in the previous subsection to sGB gravity.
The isolated NS solutions for the sGB metric corrections were previously derived in~\cite{Saffer:2019hqn}. The scalar field is generated at $\mathcal{O}(\alpha)$, which sources the metric correction at $\mathcal{O}(\alpha^2)$.  We here state relevant results from that work. 

\subsubsection{Scalar Field at $\mathcal{O}(\epsilon^0,\alpha^1)$}
\label{subsubsec:eps0_alpha1}

We begin with the expansion of the scalar field as presented in Eq.~\eqref{eq:varphiExpand}.  Recall that due to shift symmetry, we may shift our scalar field by $\varphi \rightarrow \varphi - \varphi_0$ to obtain $\varphi \sim \order{\alpha}$.  This gives us Eq.~\eqref{eq:field_FE} as a pure function of $\alpha$.  Therefore, all curvature terms will be constructed with the results found in Sec.~\ref{subsec:GR}.  The field equations for the scalar field leads to the differential equation
\begin{align}
\label{eq:interiorFieldEqforField}
    \frac{d^2 \varphi_{01}}{dr^2} &=\frac{2 \left[ m_0-r+2 \pi r^3 (\rho_{00} - p_{00})\right]}{r\left(r-2m_0\right)} \diff{\varphi_{01}} \nonumber \\
    & \quad + \frac{128 \pi r^3 \left( m_0 + 2 \pi p_{00} r^3\right)\rho_{00} - 48 m_0^2}{r^5 \left( r-2m_0 \right)}\,.
\end{align}

For the interior solution, we solve numerically the above equation with the solutions to all background terms obtained at GR order. The initial conditions for our integration are found as in GR by taking a Taylor expansion of Eq.~\eqref{eq:interiorFieldEqforField} about $r=0$ and assigning some small initial radius $r_0 \ll R_0$. This allows us to obtain the interior solution up to a constant that corresponds to a homogeneous solution to Eq.~\eqref{eq:interiorFieldEqforField} and can be found by matching the interior solution to the exterior solution (described below) at the stellar surface. 

The exterior solution may be found by the limit $p_{00} \rightarrow 0$, $\rho_{00} \rightarrow 0$, and $m_0 \rightarrow M_0$.  Such exterior scalar field may be solved analytically and found to be
\begin{align}
\label{eq:ExteriorField1}
    \varphi^{\rm ext}_{01} &= \frac{C_1}{2\,M_0} \ln \left(1-\frac{2\,M_0}{r} \right) + \frac{1}{M_0^2} \ln \left( 1 - \frac{2\,M_0}{r}\right) \nonumber \\
    &\quad + \frac{2}{r}\left( \frac{1}{M_0} + \frac{1}{r} + \frac{4\,M_0}{3\,r^2}\right) + C_2\,.
\end{align}
Requiring a vanishing scalar field at spatial infinity allows us to set $C_2 = 0$. Expanding Eq.~\eqref{eq:ExteriorField1} as $r \rightarrow \infty$, we find
\begin{equation}
    \varphi^{\rm ext}_{01} = -\frac{C_1}{r} - \frac{M_0\,C_1}{r^2} - \frac{4\,M_0^2\,C_1}{3\,r^3} + \order{\frac{1}{r^4}}\,.
\end{equation}
In this limit, we find that $C_1$ represents a scalar monopole charge for the NS (normalized by $\alpha$).
However, it has been shown that such a charge does not exist for NS in sGB theory~\cite{Yagi:2015oca}.  Therefore, we are justified in setting $C_1=0$ here and express our exterior scalar field as
\begin{equation}
        \varphi^{\rm ext}_{01} = \frac{1}{M_0^2} \ln \left( 1 - \frac{2\,M_0}{r}\right) + \frac{2}{r}\left( \frac{1}{M_0} + \frac{1}{r} + \frac{4\,M_0}{3\,r^2}\right)\,. 
\end{equation}

\subsubsection{Metric at $\mathcal{O}(\epsilon^0,\alpha^2)$}
\label{subsubsec:eps0_alpha2}

With our scalar field in hand, we may move onto the metric terms.  We may see that Eq.~\eqref{eq:metric_FE} is of $\order{\alpha^2}$, and thus all background terms refer to solutions already found. The interior solutions can be solved for numerically, but we must take care in some instances.  First, we must define the perturbation to the density $\rho_{02}$.
If we allow the total density to be written as $\rho_{00} + \alpha^2 \rho_{02} = \rho \left(p_{00} + \alpha^2 p_{02}\right)$ where $\rho(p)$ is a functional representing our equation of state, we may Taylor expand about small $\alpha$ to recover our perturbation
\begin{equation}
\label{eq:rho_2}
    \rho_{02} = p_{02} \left(\frac{d \rho_{00}}{dr}\right) \left( \frac{d \, p_{00}}{dr}\right)^{-1}\,.
\end{equation}
Next, we find a modified TOV equation 
\begin{align}
\label{eq:NewTOV}
    \diff{p_{02}} &= -\frac{1}{2r\left(r-2m_0\right)} \left[ r \left( p_{00} + \rho_{00}\right)\left( r-2m_0 \right) \diff{\tau_{02}} \right. \nonumber \\
    &\left. \quad - 8 \left( p_{02} + \rho_{02}\right)\left(p_{00} \pi r^3 + \frac{m_0}{4} \right)\right]\,,
\end{align}
which can be solved for simultaneously with the metric components (Eq.~\eqref{eq:sigma02}) to find a new boundary to the star.  We define a new boundary $R$, where the total pressure $p_{00} + \alpha^2 p_{02}$ decreases by a factor of $10^{-11}$ from the initial pressure.  We choose the initial condition by assuming the EoS is unaffected by the sGB correction.  That is, we set $\rho_{02}(r_0) = p_{02}(r_0)=0$.  Once an initial density is chosen, we find a corresponding initial pressure via the EoS, and continue the integration from there. The new radius is given by $R=R_0 + \alpha^2 R_2$ with $p_{00}(R) + \alpha^2 p_{02}(R) =0$, which can be solved order by order to yield $R_2 = -p_{02}(R_0)/p_{00}'(R_0)$. 
With this in mind, we may solve for the interior solution of the NS numerically using the equations presented in~\cite{Saffer:2019hqn} along with Eq.~\eqref{eq:NewTOV}.  The equations for the metric functions $(\tau_{02},\sigma_{02})$ can be solved for by looking at the $(t,t)$ and $(r,r)$ components of the field equations at $\order{\epsilon^0, \alpha^2}$,
\begin{widetext}
 \begin{subequations}
 \begin{align}
\label{eq:sigma02}
\diff{\sigma_{02}} &= \frac{1}{r^7 \left( r - 2 m_0\right)} \left[ 4 \pi r^8 \left( r - 2 m_0 \right) \left( \diff{\varphi_{01}}\right)^2 + 512 \pi r^4 \left( \frac{5 m_0^2}{4} - \left( \pi p_{00} r^3 + 2 \pi \rho_{00} r^3 + \frac{3}{4} r \right) m_0 + \pi \rho_{00} r^4 \right) \left( \diff{\varphi_{01}} \right) \right. \nonumber \\
&\left. \quad + 8 \pi \rho_{00} \sigma_{02} r^9 + 32768 \pi^3 p_{00} \rho_{00} m_0 r^6 + 8 \pi \rho_2 r^9 + 16384 \pi^2 m_0^2 \rho_{00} r^3 - \sigma_{02} r^7 - 6144 \pi m_0^3 \right]\,, \\
\label{eq:tau02}
\diff{\tau_{02}} &= \frac{1}{r^3 \left( r - 2 m_0\right)} \left[ 4 \pi r^4 \left( r - 2 m_0\right) \left(\diff{\varphi_{01}}\right)^2 + 512 \pi \left( \pi p_{00} r^3 + \frac{m}{4} \right) \left( r - 3 m_0 \right) \left( \diff{\varphi_{01}}\right) \right. \nonumber \\
&\left. \quad + 8 \pi p_{00} \sigma_{02} r^5 + 8 \pi p_2 r^5 + \sigma_{02} r^3 \right] \,.
 \end{align}
 \end{subequations}

Externally, the metric components may be solved for analytically as 
\begin{subequations}
\begin{align}
\label{eq:tau02Solution}
\tau_{02}^\mathrm{ext} &= \left( 1 - \frac{2\,M_0}{r}\right)^{-1} M_0^{-4} \left[ - 24\pi \left(1-\frac{7}{3}\frac{M_0}{r}\right) \ln \left( 1-\frac{
2M_0}{r}\right) \right. \nonumber \\
&\left. \quad - \frac{M_0}{r} \left( 48 \pi -64\pi \frac{M_0}{r} -48 \pi \frac{M_0^2}{r^2} + \left( C_m r^3 - \frac{160 \pi}{3}\right) \frac{M_0^3}{r^3} - \frac{352\pi}{5} \frac{M_0^4}{r^4} - \frac{512 \pi}{5} \frac{M_0^5}{r^5} + \frac{1280\pi}{3} \frac{M_0^6}{r^6}\right) \right]\,,\\
\label{eq:sigma02Solution}
\sigma_{02}^\mathrm{ext} &= \left( 1 - \frac{2\,M_0}{r}\right)^{-1} M_0^{-4} \left(\frac{M_0}{r}\right)\left[ -8\pi \ln \left(1- \frac{2M_0}{r} \right) \right. \nonumber \\
& \left. \quad - \frac{M_0}{r} \left( 16\pi + 16 \pi \frac{M_0}{r} + \left(\frac{64\pi}{3} - C_m r^3 \right)\frac{M_0^2}{r^2} + 32 \pi \frac{M_0^3}{r^3} + \frac{256\pi}{5}\frac{M_0^4}{r^4} - \frac{5888\pi}{3} \frac{M_0^5}{r^5}\right)\right]\,,
\end{align}
\end{subequations}
\end{widetext}
where $C_m$ is an integration constant and we have already ensured that the limit for the metric perturbations as $r\rightarrow \infty$ vanishes.  We also note that in the limit of $r \rightarrow \infty$, the full metric component at $\order{\epsilon^0,\alpha^2}$ becomes
\begin{equation}
\label{eq:Newgtt}
    g_{tt} = -\left(1 - \frac{2M _0+ \alpha^2 C_m}{r}\right) + \order{\frac{1}{r^2}} \,.
\end{equation}
From Eq.~\eqref{eq:Newgtt}, we see that the term we call $C_m$ is a correction to the mass.  We may therefore allow us to redefine our mass in terms of the $M_0$ from the GR contribution as well as the correction from sGB gravity as 
\begin{equation}
\label{eq:mass}
    M= M_0 + \alpha^2\frac{C_m}{2}\,,
\end{equation}
which presents our metric in the limit $r\rightarrow \infty$ as the familiar $g_{tt} = - \left( 1 - 2M/r \right)$.  Note that upon observation of the NS mass, the measured value will be $M$.

Similar to the GR case in Sec.~\ref{sec:background}, we determine the integration constants via matching the interior and exterior solutions at the background surface $R_0$ (contribution from the correction to the radius enters at higher order).  We first make the assumption that the EoS itself is unaffected by the sGB parameter at the center of the star.  Therefore our initial conditions will remain identical to the GR conditions at $r=r_0$ (i.e. $\rho_{02}(r_0)=p_{02}(r_0)=0$).  We may then solve the interior equations in Eqs.~\eqref{eq:rho_2} and~\eqref{eq:sigma02}. We then match the $\sigma_{02}$ interior with that in the exterior in Eq.~\eqref{eq:sigma02Solution} at $R_0$ to determine $C_m$. $\tau_{02}$ is determined by first choosing a trial initial condition and solve Eq.~\eqref{eq:tau02} in the interior region. We add to this a constant (corresponding to a homogeneous solution) and determine it by matching the solution to the exterior one in Eq.~\eqref{eq:tau02Solution} at the surface.

\subsection{Scalar-Gauss-Bonnet Corrections: Tidal Perturbation}
\label{subsec:LoveCorrection}
The final step of our calculation is to solve for the tidal perturbation to the scalar field and metric functions in sGB gravity.  We begin with the scalar field and make use of Eq.~\eqref{eq:field_FE} and the results of Sec.~\ref{subsec:GR} to find our solution.  Recall that the metric only contains corrections at $\order{\epsilon,\alpha^0}$ and $\order{\epsilon,\alpha^2}$.  Since $\varphi \sim \order{\alpha}$, we will only need the GR contributions to the metric in the scalar field equation.  Therefore, our correction to the scalar field will be $\order{\epsilon,\alpha}$.  The metric will be corrected at $\order{\epsilon,\alpha^2}$ as previously shown.

\subsubsection{Scalar Field at $\order{\epsilon,\alpha}$}
\label{subsubsec:ScalarLoveCorrection}

At $\order{\epsilon,\alpha}$, the field equation for $\varphi_{11}$ is given by

\begin{widetext}
 \begin{align}
 \label{eq:varphi11BigEqn}
     \frac{d^2 \varphi_{11}}{dr^2} &+ \left[\frac{1}{2}\left( \diff{\tau_{00}}-  \diff{\sigma_{00}} \right) + \frac{2}{r}\right]  \diff{\varphi_{11}} - \frac{6 e^{\sigma_{00}}}{r^2} \varphi_{11} = \frac{1}{2r^3} \left\{ 12 r \left[ \left( -\frac{2r}{3} \diff{K_{10}} - \frac{4\tau_{10}}{3}\right)\frac{d^2 \tau_{00}}{dr^2}  - \frac{2}{3} \frac{d^2 \tau_{10}}{dr^2} \right. \right. \nonumber \\ 
		&\left. \left. - \frac{2}{3r} \diff{\tau_{00}} \frac{d^2 K_{10}}{dr^2} + \left( -\frac{r}{3}\diff{K_{10}} - \frac{2}{3}\tau_{10}\right) \left( \diff{\tau_{00}}\right)^2 + \left(\left( r \diff{K_{10}}+2 \tau_{10}\right)\diff{\sigma_{00}} - \frac{4}{3}\diff{K_{10}}- \frac{5}{3}\diff{\tau_{10}} \right) \diff{\tau_{00}} \right. \right. \nonumber \\
		&\left. \left. +\diff{\tau_{10}} \diff{\sigma_{00}} \right]e^{-\sigma_{00}} + 16 r \left( K_{10} + \frac{\tau_{10}}{2}\right)  \frac{d^2 \tau_{00}}{dr^2} + 8r \frac{d^2 \tau_{10}}{dr^2}-2 r^3 \tau_{10}  \frac{d^2 \varphi_{01}}{dr^2} \right. \nonumber \\
		& \left. + 8 r \left( K_{10} + \frac{\tau_{10}}{2}\right) \left( \diff{\tau_{00}}\right)^2 + \left[ 12 r \diff{\tau_{10}}- 8 r \left( K_{10} + \frac{\tau_{10}}{2}\right)\diff{\sigma_{00}} - r^3 \tau_{10} \diff{\varphi_{01}}  -24 \tau_{10}\right] \diff{\tau_{00}} \right. \nonumber \\
		&\left.  + \left(r^3 \tau_{10} \diff{\varphi_{01}} - 4 r \diff{\tau_{10}}  - 24 \tau_{10}\right)\diff{\sigma_{00}} - 2r^2 \left(r \diff{K_{10}} + r \diff{\tau_{10}} + 2 \tau_{10}\right) \diff{\varphi_{01}} \right\}\,.
 \end{align}
\end{widetext}
For solving Eq.~\eqref{eq:varphi11BigEqn} in the interior of the star, we follow a procedure laid out in~\cite{Yagi:2013mbt} for dynamical Chern-Simons gravity.  First, we solve Eq.~\eqref{eq:varphi11BigEqn} with arbitrary initial conditions such that $\varphi(r)$ and $\varphi'(r) $ are regular at the center of the star.  This will give us a particular solution $\varphi^{\rm part}_{11}$.  Next, we solve Eq.~\eqref{eq:varphi11BigEqn} assuming the source vanishes.  Again, with arbitrary initial conditions that ensure regularity at the center of the star, we recover a homogeneous solution $\varphi^{\rm homo}_{11}$.  Our full interior solution for the scalar field may then be written
\begin{equation}
    \label{eq:FieldInterior11Temp}
    \varphi_{11}^{\rm int} = \varphi^{\rm part}_{11} + C_h \,\varphi^{\rm homo}_{11}\,,
\end{equation}
where $C_h$ is a constant to be matched at the boundary of the star with the exterior solution, which we will solve for now.

Given the complication of Eq.~\eqref{eq:varphi11BigEqn}, instead of finding the exterior  solution analytically, we make the ansatz that the scalar field is given by a polynomial series through the Taylor expansion about $r=\infty$ as 
\begin{equation}
\label{eq:FieldAnsatz}
\varphi^{\rm ext}_{11} = \sum_{k=0} \phi_{k} r^{2-k}\,.
\end{equation}
We can solve for each coefficient order by order in $r$ using Eq.~\eqref{eq:FieldAnsatz} with Eq.~\eqref{eq:varphi11BigEqn} and find only two unknown constants, $\phi_0$ and $\phi_5$.
Following~\cite{Pani:2014jra,Cardoso:2017cfl} in the case of scalar-tensor theories and Chern-Simons gravity, we set the scalar tidal field to vanish. Namely, we require the scalar field to be finite at $r \to \infty$, which leads us to $\phi_0=0$.  Therefore, we find our exterior solution to be approximately
\begin{align}
\label{eq:FieldExterior11Temp}
    \varphi_{11}^{\rm ext} &= -\frac{24 c_2}{M_0 r} - \frac{24 c_2}{r^2} + \frac{\phi_5}{r^3} - \frac{128 c_2 M_0^2 - 3 M_0 \phi_5}{r^4} \nonumber \\ 
    & - \frac{16}{7} \left( \frac{128 c_2 M_0^3 - 3 M_0^2 \phi_5}{r^5} \right) \nonumber \\
    & + \frac{4}{105} \left( \frac{504 c_1 M_0^4 - 16000 c_2 M_0^4 + 375 M_0^3 \phi_5}{r^6} \right) \nonumber \\
    & + \frac{8}{105} \left( \frac{812 c_1 M_0^5 - 16000 c_2 M_0^5 + 375 M_0^4 \phi_5}{r^7} \right) \nonumber \\
    & + \frac{8}{2625} \left( \frac{50276 c_1 M_0^6 - 784000 c_2 M_0^6 + 18375 M_0^5 \phi_5}{r^8} \right)\,,
\end{align}
where we have kept up to $\order{1/r^8}$ which we found ensures that our series for $\varphi_{11}^{\rm ext}$ converges.

We may now match Eqs.~\eqref{eq:FieldInterior11Temp} and~\eqref{eq:FieldExterior11Temp} (and also their first derivatives) at the stellar boundary, $R_0$.  This method allows us to find the solutions for the interior and exterior numerically, and we are left with constants $C_h$ and $\phi_5$.  Due to the construction of the interior solution, the arbitrariness of the initial conditions is absorbed into the constant term in front of the homogeneous solution.  Therefore, as long as we ensure regularity at the center of the star, the exterior solution will remain independent of the interior conditions chosen.

\subsubsection{Metric at $\order{\epsilon,\alpha^2}$}
\label{subsubsec:MetricLoveCorrection}
We finally study the tidal, sGB correction to the metric at $\order{\epsilon,\alpha^2}$.
We may solve for the metric equations as well as the density and pressure perturbations via the field equations in  Eq.~\eqref{eq:metric_FE} and conservation identities.  Keeping both sides of the equation to $\order{\epsilon,\alpha^2}$ we mimic the approach from Sec.~\ref{subsec:GR}.  Similar to the GR case, conservation of the matter stress energy tensor yields the relations similar to Eq.~\eqref{eq:SetConserve1} for the new perturbations $(p_{12},\rho_{12})$:
\allowdisplaybreaks
\begin{subequations}
\label{eq:SetConserve2}
\begin{align}
    p_{12} &= - \frac{\sqrt{5 \pi}}{4 \pi} \left\{\left( p_{02} + \rho_{02}\right)\tau_{10} + \left(p_{00} + \rho_{00}\right) \tau_{12}\right]\,,\\
    \rho_{12} &= \frac{1}{4 \pi p_{00} r^3 + m_0} \left[ \frac{r}{2} \left( p_{10} + \rho_{10}\right) \left( r - 2m_0\right) \diff{\tau_{02}} \right. \nonumber \\
    &\left. + \frac{\sqrt{5 \pi}}{\pi} \left[ \frac{4}{r} \left(r - 2m_0 \right) \left( \tau_{10} \diff{p_{02}} \right. \right. \right. \nonumber \\
    &\left. \left. \left.+ \tau_{10} \diff{\rho_{02}} +\tau_{12} \left( \diff{p_{00}} + \diff{\rho_{00}}\right)\right)\right]\right\}\,.
\end{align}
\end{subequations}
Subtracting the $(\phi,\phi)$ component from the $(\theta,\theta)$ component provides one relation between two metric components, while the $(r,\theta)$ component provides another:
\begin{widetext}
\begin{subequations}
\begin{align}
    \sigma_{12} &= -\tau_{12} + 32\pi e^{-\sigma_{00}}\left\{ 2 \tau_{10} \left( \frac{d^2 \varphi_{01}}{dr^2} \right) + 2 \varphi_{11} \left( \frac{d^2 \tau_{00}}{dr^2}\right) + \varphi_{11} \left( \frac{d\,\tau_{00}}{dr}\right)^2 \right. \nonumber \\
    &\left. \quad + \left[ \left(\frac{d\,\varphi_{01}}{dr}\right)\sigma_{10} - \varphi_{11} \left(\frac{d\, \sigma_{00}}{dr} \right) \right]\left( \frac{d\,\tau_{00}}{dr}\right)- \tau_{10}\left(\frac{d \, \sigma_{00}}{dr}\right)\left(\frac{d\, \varphi_{01}}{dr}\right)\right\}\,, \\
    \frac{d\,K_{12}}{dr} &= \frac{1}{2r^2} \left\{ 64 \pi e^{-\sigma_{00}} \left[\left(r \left( 4\tau_{10}+ r\frac{d\,K_{10}}{dr} \right)\frac{d\,\varphi_{01}}{dr} - 2r \frac{d\,\varphi_{11}}{dr}-2 \varphi_{11}\right) \frac{d\, \tau_{00}}{dr}+2\left(r \frac{d\,\tau_{10}}{dr} - \tau_{10}\right)\frac{d\,\varphi_{01}}{dr}\right] \right. \nonumber \\
    &\left. \quad -32r\left[ \frac{r}{32}\left(\tau_{12} - \sigma_{12}\right)\frac{d\,\tau_{00}}{dr} + \pi r \varphi_{11} \frac{d\,\varphi_{01}}{dr} + \frac{r}{16} \tau_{10} \frac{d \, \tau_{02}}{dr} + \frac{r}{16} \frac{d\, \tau_{12}}{dr} - \frac{\tau_{12}+\sigma_{12}}{16}\right]\right\}\,.
\end{align}
\end{subequations}
\end{widetext}
Armed with these relations, we may express the field equations $(t,t) - (r,r)$ in terms of the metric component $\tau_{12}$ and previously solved quantities.  This results in the differential equations
\begin{equation}
    \label{eq:tau12DiffEq}
    \frac{d^2 \, \tau_{12}}{dr^2} + B_{\rm \tau} \frac{d\,\tau_{12}}{dr} + C_{\rm \tau} \tau_{12} = S_{\tau}\,,
\end{equation}
where the coefficients $B_\tau$ and $C_\tau$ are given by
\begin{widetext}
\begin{align}
    B_\tau &= \frac{4\pi r^3\left(p_{00} - \rho_{00} \right) + 2 \left( r-m_0 \right)}{r \left( r-2m_0\right)}\,, \\
    C_\tau &= -\frac{4\pi r^2}{4 \pi r^3 p_{00} +m_0} \left( \diff{\rho_{00}}\right) - \frac{4m_0^2 + \left( 104 \pi p_{00} r^3 + 40 \pi r^3 \rho_{00} - 12 r\right)m_0+64 \pi^2 p_{00}^2 r^6-36 \pi p_{00} r^4 - 20 \pi r^4 \rho_{00} + 6 r^2}{r^2 \left(r-2m_0 \right)^2}\,, 
\end{align}
\end{widetext}

Let us now discuss the source term $S_{\tau}$ in Eq.~\eqref{eq:tau12DiffEq}.  The full expression is rather lengthy, which we provide in a supplementary Mathematica notebook~\cite{saffer_2021}.
However, we found that the tidal Love number from the solution to Eq.~\eqref{eq:tau12DiffEq} with the full source expression suffers from some ambiguity that we discuss in Sec.~\ref{sec:ambiguity}. To overcome this, we consider working within a post-Minkowskian (or post-Newtonian) approximation, where we assume $m \ll r$ and expand the differential equation about $m=0$. The leading, $\mathcal{O}(m^0)$ corresponds to the Newtonian contribution, while higher order terms correspond to post-Newtonian contributions. 
We found that the ambiguity is absent if we only keep the source term at $\mathcal{O}(m^0)$, which is given by
\begin{equation}
\label{eq:source_leading}
S_{\tau}^{(0)} = -\frac{2 \pi \tau_{10} r^2}{m_0^2} \left[\sigma_{02} \left(\diff{\rho_{00}}\right) - 2m_0 \left(\diff{\rho_{02}} \right) \right] \,.
\end{equation}
Thus, we only work to this order in the source and compute the Love number. In Appendix~\ref{app:Convergence}, we show that $\mathcal{O}(m^0)$ indeed gives us the dominant contribution to the source term by solving Eq.~\eqref{eq:tau12DiffEq} with the full source term and comparing it with the case with the leading source term in Eq.~\eqref{eq:source_leading}.

We solve for $\tau_{12}$ similar to Eq.~\eqref{eq:FieldInterior11Temp} by solving the homogeneous and particular parts of Eq.~\eqref{eq:tau12DiffEq} separately ensuring that the metric function is regular at the center of the star.  The full interior solution will take the form
\begin{equation}
    \label{eq:MetricInterior11Temp}
    \tau_{12}^{\rm int} = \tau^{\rm part}_{12} + D_h \,\tau^{\rm homo}_{12}\,,
\end{equation}
where $D_h$ is a constant which must be matched to the exterior solution at the boundary of the star $R_0$.
For the exterior solution, we again make use of a polynomial series ansatz
\begin{equation}
    \label{eq:LoveMetricExpansionAnsatz}
    \tau^{\rm ext}_{12} = \sum_{k=0} \tau_{k} r^{2-k}\,,
\end{equation}
valid  to $\order{r^{-9}}$ to ensure our result converges to a solution, 
and determine each coefficient order by order in $r$.
For example, the exterior solution for the leading order source term at $\mathcal{O}(m^0)$ is
\begin{align}
\label{eq:tau12Order0Expansion}
    \tau_{12}^{\rm ext} &= \tau_0 r^2 + - 2 M_0 \tau_0 r + \tau_5 \left(\frac{1}{r^3} + \frac{3M_0}{r^4} + \frac{50M_0^2}{7r^5} \right. \nonumber \\
    &\left. \quad + \frac{110 M_0^3}{7r^6} + \frac{100 M_0^4}{3 r^7} + \frac{208 M_0^5}{3 r^8} + \frac{1568 M_0^6}{11 r^9}\right)\,.
\end{align}

\subsection{Love Number and Tidal Deformability}
\label{sec:LoveAndTidal}

We are now ready to define and explain how to compute the tidal Love number and deformability, and some ambiguity associated to it.

\subsubsection{Definition}
 
The tidal deformability $\lambda$ is defined as~\cite{Hinderer:2007mb}
\begin{equation}
Q_{ij} = - \lambda \mathcal{E}_{ij}\,,
\end{equation}
where $\mathcal{E}_{ij}$ is the quadrupolar external tidal field while $Q_{ij}$ is the tidally-induced quadrupole moment of a neutron star. The former (latter) can be read off from the $\ell =2$ part of the $r^2$ ($r^{-3}$) piece in the asymptotic behavior of the metric function $\tau$ at infinity. It is convenient to study the dimensionless tidal deformability:
\begin{equation}
\label{eq:Lambda_def}
\Lambda \equiv \frac{\lambda}{M^5}\,.
\end{equation}
This quantity is related to the tidal Love number or the second apsidal constant as $k_2 \equiv (3/2) \Lambda C^5$, where $C \equiv M/R$ is the stellar compactness. In GR, $\Lambda$ is computed from the integration constants $c_1$ (related to the quadrupole moment) and $c_2$ (related to the external tidal field) in Eq.~\eqref{eq:tau10ExteriorExpansion} as~\cite{Hinderer:2007mb}
\begin{equation}
\label{eq:lambdaBarDef}
\Lambda_0 =\frac{8\,c_1}{45\,c_2}\,.
\end{equation}

How does Eq.~\eqref{eq:lambdaBarDef} change in sGB gravity? There are two main corrections: (i) $c_1$ in Eq.~\eqref{eq:lambdaBarDef} and (ii) $M$ in Eq.~\eqref{eq:Lambda_def}. The former is corrected to $c_1 + \alpha^2 \delta c_1$ with $\delta c_1 = \frac{5}{8M_0^3} \tau_5$, while the latter is corrected to $M_0 + \alpha^2 M_2$ where $M_2 = \frac{C_m}{2}$ from Eq.~\eqref{eq:mass}. $c_2$ is uncorrected since we have set the sGB correction to the tidal field to zero (which corresponds to absorbing the tidal field correction to the GR contribution). Putting these together, we find the dimensionless tidal deformability in sGB gravity as
\begin{eqnarray}
\label{eq:CorrectedLove}
\Lambda &=& \frac{8\,\left(c_1 + \alpha^2 \delta c_1\right)}{45\,c_2} \left( 1 + \alpha^2\frac{M_2}{M_0} \right)^{-5}\,.
\end{eqnarray}

\subsubsection{Ambiguity in Love}
\label{sec:ambiguity}

We now comment on the potential ambiguity in the definition of the Love number or tidal deformability~\cite{Pani:2015hfa}\footnote{See~\cite{Gralla:2017djj} for another type of ambiguity in the Love number.} in sGB gravity. To compute the Love number, we extract the tidal field strength from the coefficient of the growing mode (whose leading order is $r^2$) in the asymptotic behavior of $\tau$, while we determine the quadrupole moment from the coefficient of the decaying mode (whose leading order is $r^{-3}$). However, there is no unique way to separate these two modes \textit{a priori}. 

Let us study the asymptotic behavior of $\tau_{10}$ in GR in Eq.~\eqref{eq:tau10ExteriorExpansion} as an example. Here $c_1$ ($c_2$) is the coefficient of the decaying (growing) mode. If we now shift $c_1$ as $c_1 = \gamma_1 + c_2 \hat{\gamma}_1$ for constants $\gamma_1$ and $\hat \gamma_1$ and absorb terms proportional to $c_2 \hat \gamma_1$ to the growing mode, the coefficient of the decaying mode now changes to $\gamma_1$. Namely, one can always absorb a part of the decaying mode to the growing mode and this is why there is no unique split of the growing and decaying modes unless we specify how to do so.
One way to alleviate this issue is to perform an analytical continuation in the number of spacetime dimensions $d$.  This method is discussed in~\cite{Pani:2015hfa,Kol:2011vg} and shows that by applying this technique in GR, one may obtain separate solutions for the growing and decaying modes, corresponding to $c_2$ and $c_1$ in our notation.

A practically simpler method of identifying the growing/decaying modes was proposed in~\cite{Pani:2015hfa}\footnote{See~\cite{Landry:2015zfa} for an alternative prescription.}. The prescription gave there was to find the constant $c_1$ such that the growing mode only contains \emph{finite} number of terms when expanded about $r=\infty$. This is indeed the case in the solution for $\tau_{10}$ in GR in Eq.~\eqref{eq:LoveExtSolution}, where the growing mode only contains terms of $\mathcal{O}(r^2)$ and $\mathcal{O}(r)$. The prescription has been shown to work when computing the Love number for slowly rotating compact objects in GR (up to quadratic order in spin for black holes and first order in spin for stars)~\cite{Pani:2015hfa}.

We here apply this prescription to sGB gravity to see whether the ambiguity exists in the calculation of the Love number. We begin by considering $\tau_{12}$ at $\mathcal{O}(m^0)$ in the post-Minkowskian expansion. The exterior solution is given in Eq.~\eqref{eq:tau12Order0Expansion}. Notice that the growing mode only contains terms proportional to $\order{r}$ and $\order{r^2}$ (similar to the GR case). This means that the growing mode only contains a finite number of terms and thus we expect one can uniquely identify the growing and decaying modes to compute the Love number. 
We found that this is no longer the case with higher order post-Minkowskian expansion and thus we focus on the leading post-Minkowskian result to avoid the ambiguity in the definition of the Love number.

\section{Astrophysical Comparison Results}
\label{sec:Astro}

In this section, we will present the results of some astrophysical studies in order to place potential bounds on the sGB correction terms.  
Ideally, one should reanalyze the data collected by LIGO/Virgo, NICER, and radio telescopes with the sGB waveform templates, pulse profiles, and timing residuals to estimate the mass, radius, etc. if one wants to use these quantities to test sGB gravity. However, the estimate of these under the GR assumption can be a good approximation for the following reason. In sGB gravity, NSs do not carry scalar charges at $\order \alpha$ which suppresses the scalar dipole radiation from a NS binary. Moreover, the exterior spacetime of a non-rotating NS is almost identical to the Schwarzschild metric in GR with the  difference entering at $\mathcal{O}(M_0^7/r^7)$~\cite{Saffer:2019hqn}. These suggest that the sGB correction to the gravitational waveform, pulse profile, and Shapiro time delay may enter at high order. For simplicity, we use the GR estimates of the NS quantities in this paper and leave a more detailed analysis for future work. In App.~\ref{app:Systematics}, we focus on the tidal deformability measurement with gravitational wave observations and compare the leading tidal effect in GR with the sGB contribution to justify our choice.

\begin{figure}
	\includegraphics[width=8.5cm]{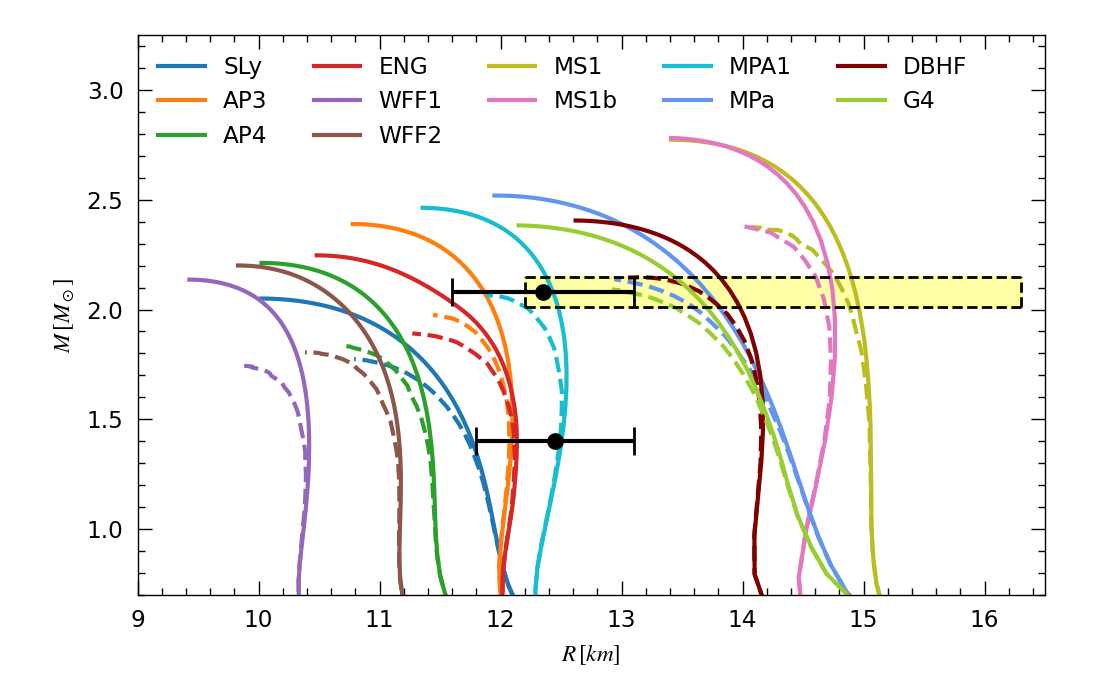}
    \caption{Similar to Fig.~\ref{fig:MRcurve_max_mass} but with $\zeta = 0.5$.
}
    \label{fig:MRcurve}
\end{figure}

Figure~\ref{fig:MRcurve} shows the mass-radius relations for GR and sGB with $\zeta = 0.5$ in a number of EoS. 
Notice that the maximum mass for each EoS in sGB gravity is smaller than the GR one, which was first found in~\cite{Saffer:2019hqn}. 
Comparing this with a measurement of $\sim 2M_\odot$ pulsars~\cite{Cromartie:2019kug,Riley:2021pdl,Miller:2021qha,Raaijmakers:2021uju}, one can constrain sGB gravity for each EoS\footnote{See Ref.~\cite{Pani:2011xm} for a similar analysis on constraining Einstein-dilaton Gauss-Bonnet gravity from the investigation of the NS maximum mass using an APR EoS.}. For example,  while a NS governed by the SLy EoS is valid in GR in terms of its maximum mass, sGB gravity with a $\zeta=0.5$ is ruled out from observations. 
Figure~\ref{fig:MRcurve_max_mass} shows a similar mass-radius relation but with the maximum value of $\zeta$ allowed for each EoS to support a NS with 2.01$M_\odot$, the lowest bound on the maximum observed NS mass provided in~\cite{Cromartie:2019kug,Fonseca:2021wxt}. Observe that, in general, the bounds are stronger for softer EoS. For all the EoS considered in this paper, the most conservative bound on sGB gravity comes from the stiffest EoS, MS1, which gives the bound $\sqrt{\alpha} < 1.29$km\footnote{This bound comes from $\zeta <1.78$, which satisfies the small coupling approximation of $\bar \zeta \ll 1$ in Eq.~\eqref{eq:bar_zeta}.}, which is comparable to other existing bounds mentioned in Sec.~\ref{subsec:sGBAction}. If we assume that the radius bounds at $1.4M_\odot$ and $2.08M_\odot$ in~\cite{Miller:2021qha} hold also in sGB gravity, MS1 is inconsistent with such measurements and the conservative bounds should come from MPA1 out of all the EoS that we consider here, which gives $\sqrt{\alpha} < 0.993$km.
A better understanding of the EoS is necessary to place limits on $\zeta$ based on mass measurements alone.

\begin{figure}
	\includegraphics[width=\linewidth]{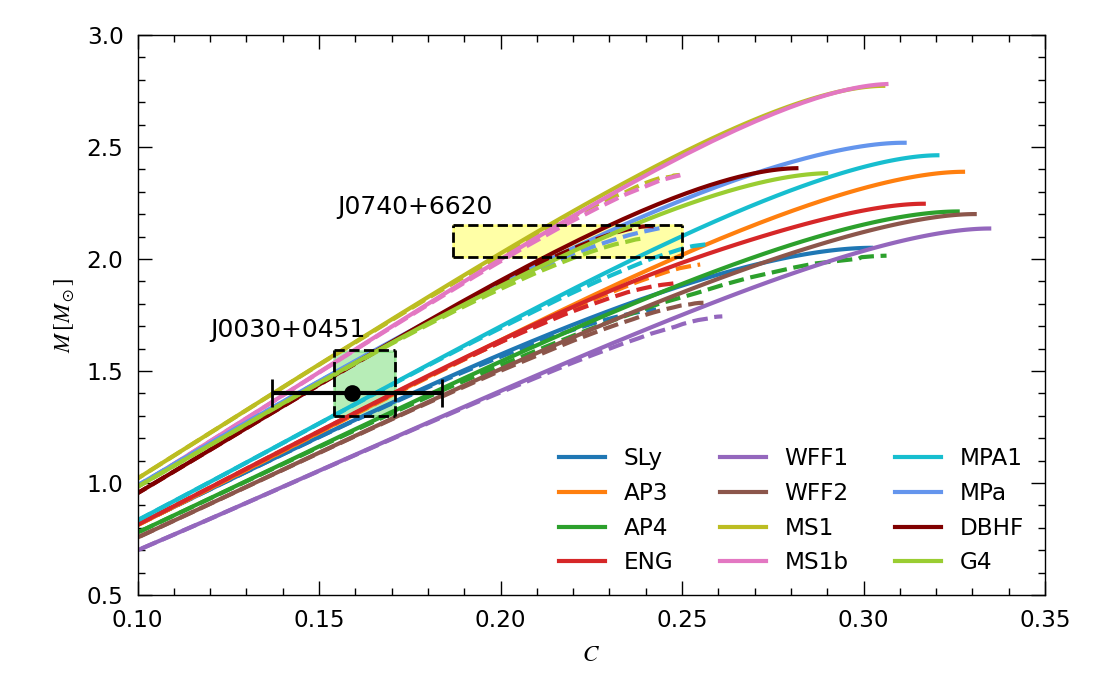}
    \caption{ Mass-compactness relations in GR (solid) and in sGB gravity with $\zeta = 0.5$ (dashed) for various EoS. We also include the mass and compactness estimates with 1-$\sigma$ errors for two NSs, J0740+6620 (yellow box)~\cite{Cromartie:2019kug,Miller:2021qha} and J0030+0451 (green box)~\cite{Miller:2019cac}, as well as recent compactness bounds for a $1.4 M_\odot$ NS (black dot with 90\% credible error bars) inferred from NICER data~\cite{Silva:2020acr}. }
    \label{fig:MCcurve}
\end{figure}

NICER has measured not only the radius but also the compactness of NSs. Figure~\ref{fig:MCcurve} shows the relation between the mass and compactness in GR and sGB with $\zeta = 0.5$, together with  constraints from the two pulsars observed by NICER. It would be difficult to use the measurement of J0030+0451 to constrain sGB gravity as the deviation from GR only becomes noticeable when the NS mass or compactness becomes relatively large. We have a better prospect of constraining the theory with J0740+6620, though the bound will depend on the choice of EoS, similar to the mass-radius case.

\begin{figure*}
	\includegraphics[width=0.9\textwidth]{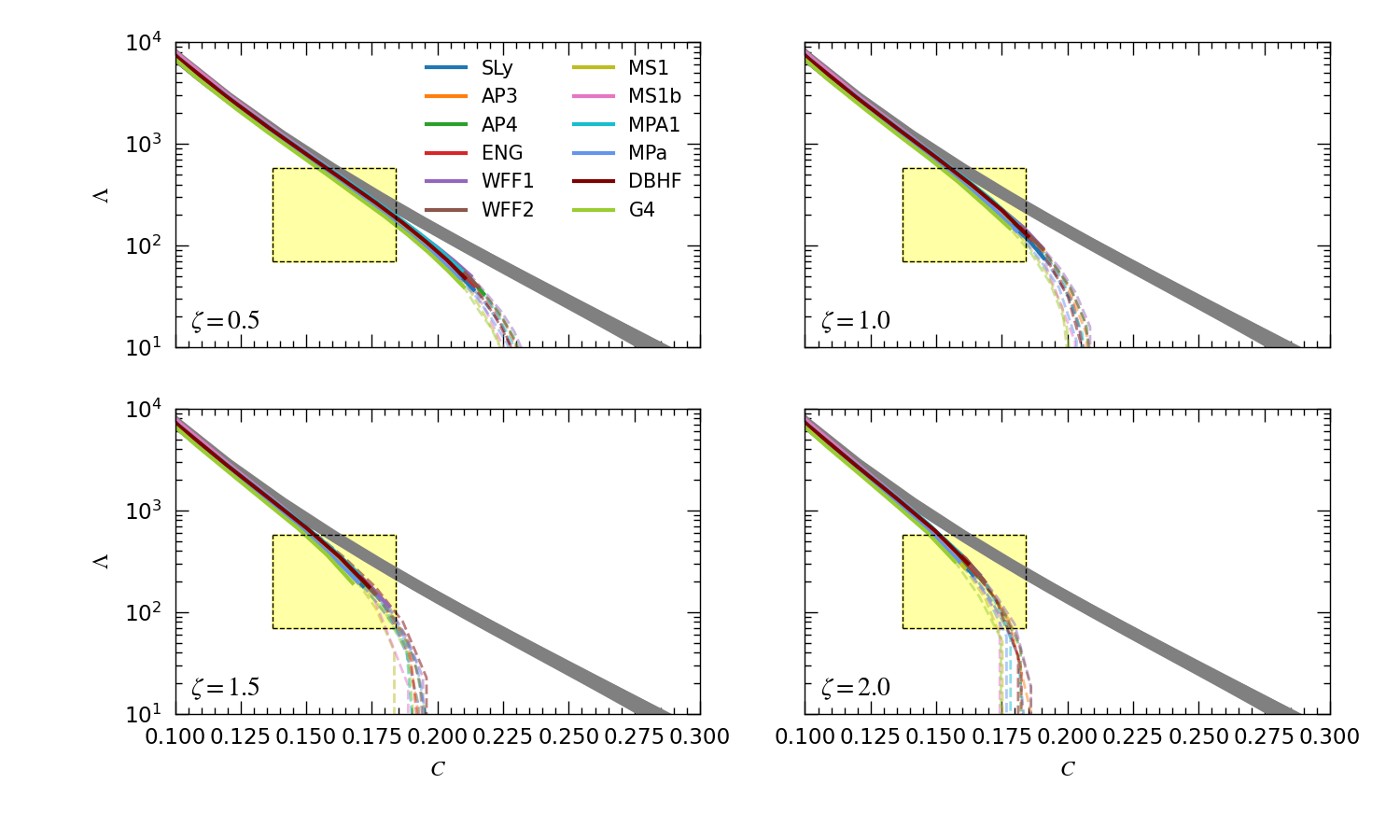}
    \caption{
    $\Lambda$--$C$ relation in GR (grey band) and sGB gravity with various $\zeta$ values (colored).
    We find that the universality tends to hold, however there is no noticeable limits which can be placed on $\zeta$ via the 90\%-credible tidal deformability measurement of GW170817~\cite{Abbott:2018exr} and the 1-$\sigma$ compactness measurement of J0030+0451~\cite{Silva:2020acr}, both at $1.4M_\odot$ (yellow box). 
    The dashed lines correspond to the continuation of the sGB solution following $\Lambda_{\rm sGB} < 0.5 \Lambda_{\rm GR}$, where we take the small coupling approximation to no longer be valid. }
    \label{fig:FullLambdaC}
\end{figure*}

One may find bounds on the theory that are less sensitive to EoS through universal relations. Here, we focus on the relation between the dimensionless tidal deformability $\Lambda$ and compactness $C$ that is known to be universal in GR~\cite{Maselli:2013mva,Yagi:2016bkt}. Figure~\ref{fig:FullLambdaC} shows the $\Lambda$--$C$ relation in GR and sGB gravity with various values of $\zeta$. Observe that the relation is still universal in sGB gravity for a fixed $\zeta$ when the sGB correction to $\Lambda$ is smaller than the GR value by 50\% (beyond this, the small coupling approximation may be invalid).
Notice also that $\Lambda$ in sGB gravity is smaller than that in GR for a fixed $C$ and the deviation from GR becomes larger as the compactness (and thus the stellar curvature) increases. To check our numerical calculation, we present in Appendix~\ref{app:CD} an analytic derivation of the $\Lambda$--$C$ relation to leading order in the post-Minkowskian approximation for constant density stars. We find a qualitatively similar behavior as for the realistic EoS case (that the sGB effect makes the tidal deformability lower and the deviation from GR increases as one increases the compactness).

To apply this universal relation to the measurement of $\Lambda$ and $C$ obtained from different system (e.g. GW170817 for the former and J0030+0451 for the latter), one needs to first convert the measurement of $\Lambda$ and $C$ at the same mass. The LIGO/Virgo Collaboration has derived a  bound on $\Lambda$ for a NS with a mass of 1.4$M_\odot$. The compactness bound from J0030+0451 for the same mass has been obtained in~\cite{Silva:2020acr}. We show these measurement errors of NSs at 1.4$M_\odot$ as yellow boxes in Fig.~\ref{fig:FullLambdaC}. For all $\zeta$ values considered in the figure, both the GR and sGB relations go through the error box, which suggests that it would be difficult to constrain sGB gravity with observations of  GW170817 and J0030+0451. This is because the stellar curvature of NSs with $1.4M_\odot$ is not large enough and a potential sGB effect is too small to be probed with a combination of observations of these astrophysical systems.

\begin{figure*}
	\includegraphics[width=8.5cm]{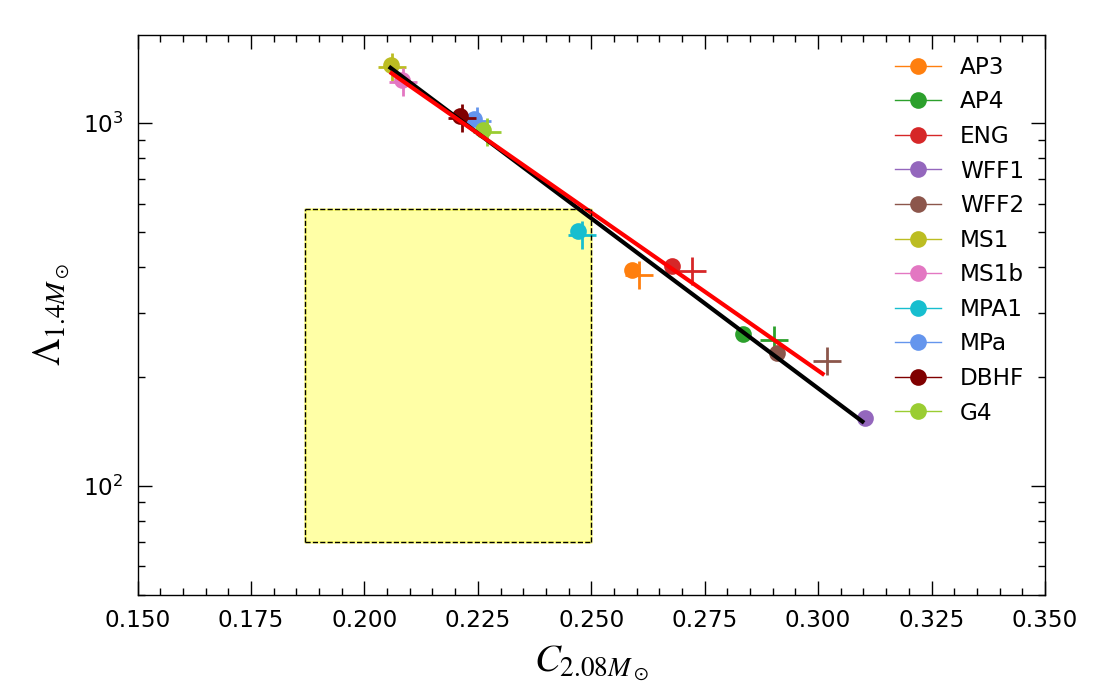}
	\includegraphics[width=8.5cm]{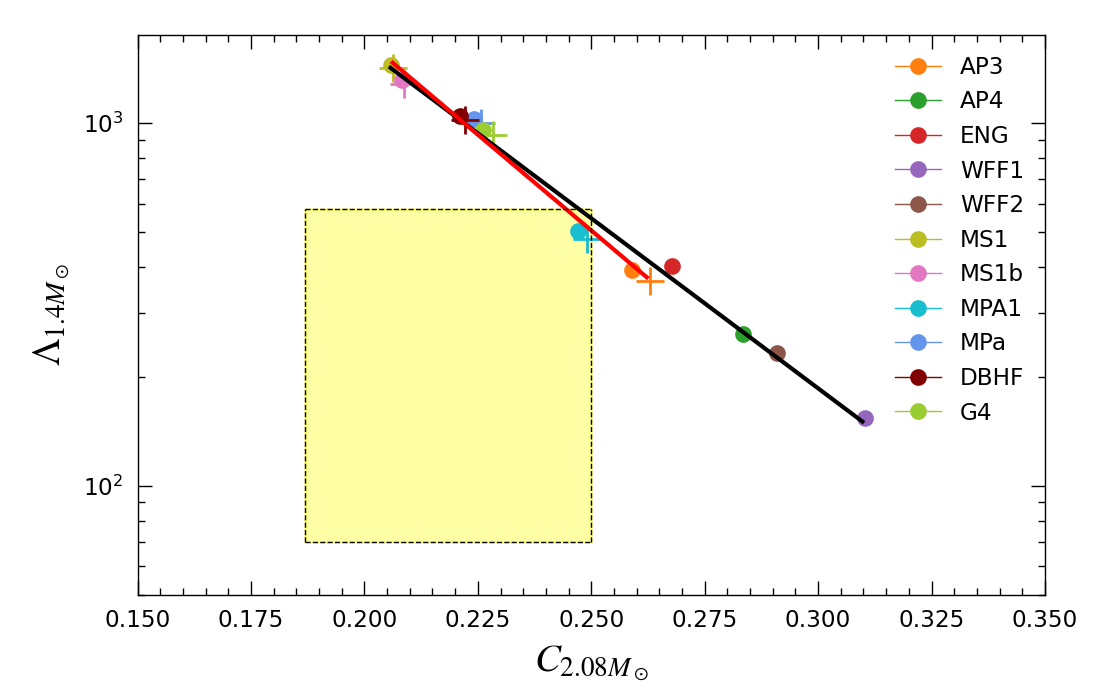}
    \caption{(Left) Relation between $\Lambda$ at $1.4M_\odot$ and $C$ at $2.08M_\odot$ in GR (dots) and in sGB gravity with $\zeta=0.1$ (+) with various EoS. We also show the fit for the relation in GR (black) and sGB gravity (red) given in Eq.~\eqref{eq:fit}. Notice the strong correlation between the two observables. The shaded region (yellow) shows the 90\% credible measurement of $\Lambda$ with GW170817 and 1-$\sigma$ measurement of $C$ with J0740+6620. (Right) Similar to the left panel but for $\zeta = 0.2$. Some of the soft EoS do not have a sGB correspondence because their maximum mass is below $2.08M_\odot$, which changes the behavior of the sGB fit with respect to the GR one from the left panel. Given that both GR and sGB relations are marginally consistent with the gravitational wave and x-ray measurements, we cannot find any meaningful bounds on sGB gravity using this relation yet, which may change if we include more EoS or if we have more accurate measurements with future observations.
}
    \label{fig:Love-C-diff-mass}
\end{figure*}

We can investigate the possibility of using universal relations in an alternate way by making use of the $\Lambda$ and $C$ relations for different masses.  We choose $1.4 M_\odot$ for the tidal deformability measured from GW170817~\cite{Abbott:2018exr} and $2.08 M_\odot$ for the compactness inferred from J0740-6620~\cite{Miller:2021qha,Riley:2021pdl}. Figure~\ref{fig:Love-C-diff-mass} presents the relation between such quantities in GR and sGB gravity with two different choices of $\zeta$, for various EoS. Notice that there is a strong correlation between $\Lambda_{1.4M_\odot}$ and $C_{2.08M_\odot}$, similar to the $\Lambda$--$C$ relation for the same NS masses in Fig.~\ref{fig:FullLambdaC}. We also show fits to the relation in each theory given by
\begin{equation}
\label{eq:fit}
\log \Lambda_{1.4M_\odot} = a_0 + a_1 C_{2.08M_\odot}\,,
\end{equation}
with the coefficients given in Table~\ref{tab:fit}. The EoS variation in the $\Lambda_{1.4M_\odot}$--$C_{2.08M_\odot}$ relation is $\sim 10\%$.

\renewcommand{\arraystretch}{1.2}
\begin{table}[tb]
\begin{centering}
\begin{tabular}{c|c|c|c}
\hline
\hline
\noalign{\smallskip}
& GR & sGB ($\zeta = 0.1$) & sGB ($\zeta = 0.2$)  \\
\hline
$a_0$ & 11.702 & 11.363 & 12.307\\
$a_1$ & -21.593 & -20.087 & -24.319\\
\noalign{\smallskip}
\hline
\hline
\end{tabular}
\end{centering}
\caption{Fitting coefficients in Eq.~\eqref{eq:fit} for the $\Lambda_{1.4M_\odot}$--$C_{2.08M_\odot}$ relation in Fig.~\ref{fig:Love-C-diff-mass}.
}
\label{tab:fit}
\end{table}

Let us now discuss whether one can place bounds on sGB gravity through the $\Lambda$ measurement of GW170817 and the $C$ measurement of J0740+6620 using the $\Lambda_{1.4M_\odot}$--$C_{2.08M_\odot}$ relation. We show the measurement errors from these observations as a yellow box in Fig.~\ref{fig:Love-C-diff-mass}. First, notice that the theoretical prediction in GR is only marginally consistent with the error box, which is due to a slight tension in these measurements that GW170817 prefers softer EoS while J0740+6620 prefers stiffer EoS. Second, notice that the relations in sGB gravity are also consistent with the measurements. As we increase $\zeta$, there is less number of EoS that can support a $2.08_{M_\odot}$ NS and it becomes more difficult to draw a robust conclusion from the universal relation with only the EoS considered in this paper. Additionally, it is difficult to determine whether the  scatter seen with the points in Fig.~\ref{fig:Love-C-diff-mass} is dominated by the EoS-variation in the relation or the difference in gravitational theories.
Thus, one needs to carry out a more detailed analysis with a significant increase in the number of EoS to see whether one can place a meaningful bound on sGB gravity from this new type of universal relations for NSs with different masses. Such a relation may provide a new way of combining different NS observations in the multimessenger astronomy era to probe strong-field gravity.

\section{Conclusion and Discussion}
\label{sec:Discussion}

In this work, we combined different NS observations to probe sGB gravity where a quadratic curvature term is present in the action. In particular, we derived a correction to the tidal deformability in this theory for the first time. Our method made use of a perturbative scheme in terms of both small tidal deformation and small sGB coupling constant. Furthermore, keeping only the leading post-Minkowskian part in the source term of the field equation at linear order in tidal deformation and quadratic order in the sGB coupling, we were able to avoid ambiguities in defining a Love number by allowing for a separation of the growing and decaying modes which is not apparent in the full solution.

We found the following main results. For NSs without tidal deformation, we found that the maximum mass of a NS decreases as one increases the sGB coupling constant. This allowed us to set an upper bound on the theory that is EoS dependent. Taking the stiffest EoS considered in this paper that gives us the most conservative bound, we derived a bound that is comparable to other existing bounds from BH observations. For tidally-deformed NSs, we found that the sGB correction to the dimensionless tidal parameter $\Lambda$ increases as one increases the NS compactness $C$. Moreover, the relation between $\Lambda$ and $C$ has been known to be EoS-insensitive in GR, and such universality is preserved in sGB gravity for a fixed dimensionless coupling constant $\zeta$, though the relation itself deviates from GR, especially at large $C$. We next applied this universal relation to astrophysical observations by LIGO/Virgo and NICER. We found that from the tidal deformability and compactness measurement of a NS at $1.4M_\odot$, it is difficult to constrain the theory via the universal relation as the deviation from GR is too small.
We also compared the $\Lambda$ and $C$ relation for different mass systems (for GW170817 and J0740+6620).  Through this avenue we found that, at the current moment, no significant bounds can be placed on the theory.  However, we find this method to be useful and worth consideration in the future as more data and observations become available.

We end by providing several avenues for future work. First, it is important to study in more detail the ambiguity in the Love number in sGB gravity. One can apply analytic continuation and see if one can unique identify the tidally-induced quadrupole moment from the asymptotic behavior of the metric. 
Second, ideally, one should reanalyze the data obtained by LIGO/Virgo and NICER with the sGB waveform template and pulse profile to estimate $\Lambda$ and $C$ without assuming GR. 
Third, we need to refine the relation between $\Lambda$ and $C$ for different mass systems presented in this paper by studying broader classes of EoS.
Lastly, it would be useful to construct a parameterized fit for the Love-C relation that includes the sGB one as an example, which is similar to what has been done in~\cite{Silva:2020acr} for the I-Love relation with dynamical Chern-Simons gravity.

\acknowledgments
We thank Tyler Gorda for initial discussions that led to this work.
K.Y. acknowledges support from NSF Grant PHY-1806776, NASA Grant 80NSSC20K0523, a Sloan Foundation Research Fellowship and the Owens Family Foundation. 
K.Y. would like to also acknowledge support by the COST Action GWverse CA16104 and JSPS KAKENHI Grants No. JP17H06358.

\appendix

\section{Convergence Test}
\label{app:Convergence}
\begin{figure}
	\includegraphics[width=\linewidth]{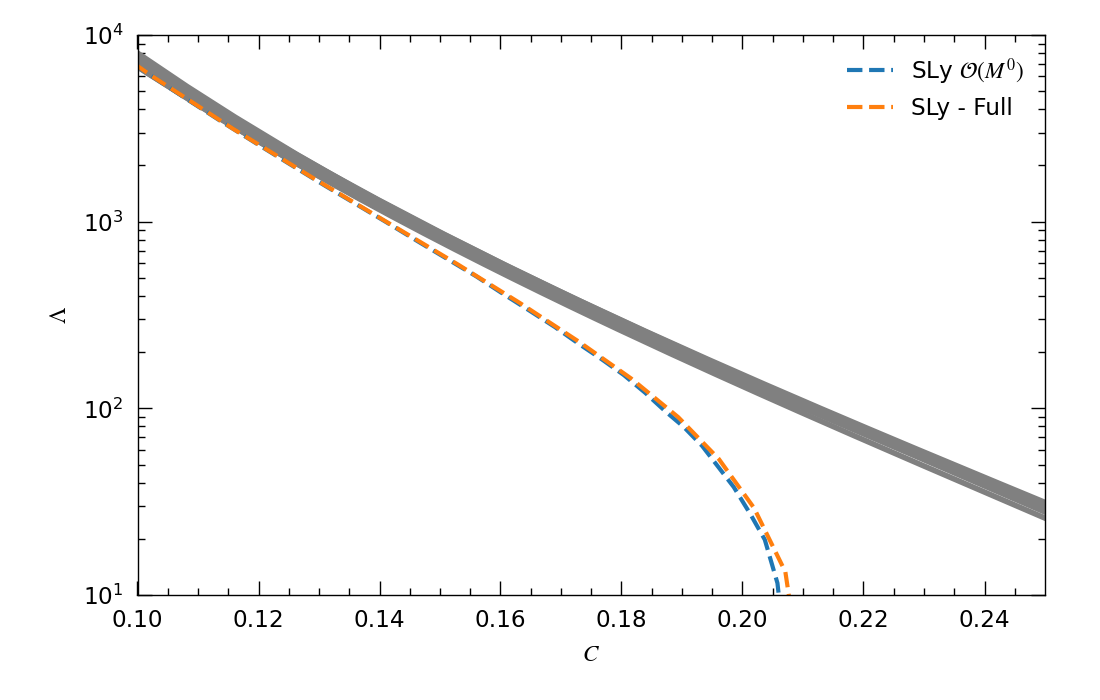}
    \caption{A check of the validity of our post-Minkowskian expansion at $\order{\epsilon,\alpha^2}$. We present the Love-C relation with the leading post-Minkowskian source term at $\order{M^0}$ and the full one in Eq.~\eqref{eq:tau12DiffEq}. Notice that the former has little deviation from the latter. Here, we use $\zeta=1.0$ for the sGB term.}
    \label{fig:Convergence}
\end{figure}

Since our approach for identifying the Love number is based on an expansion in the mass terms, we need to check if we are keeping enough terms in the source term in Eq.~\eqref{eq:tau12DiffEq} to show that our solution is converging. To achieve this, we solve Eq.~\eqref{eq:tau12DiffEq} with the full source term and compute the tidal deformability assuming that the $1/r^3$ part of $\tau_{12}$ in its asymptotic behavior contains purely the quadrupole moment contribution. That is, we do not consider our full result to be contaminated by the ambiguity discussed in Se.~\ref{sec:ambiguity}.
Figure~\ref{fig:Convergence} shows our results of this check, where we compare the tidal deformability of the leading post-Minkowskian source and the full source in Eq.~\eqref{eq:tau12DiffEq}.
We see that the leading order solution and the solution presented with the full source term are sufficiently close. This justifies that the leading post-Minkowskian contribution in the source is indeed the dominant term and provides support to our post-Minkowskian analysis that evades the ambiguity in the Love calculation.
 
\section{Estimate of Systematic Errors to Tidal Deformability Measurement}
\label{app:Systematics}
 
It is important for us to check to ensure that the systematic errors on measurements of NS observable due to the GR assumption are negligible when using them to test sGB gravity. Given that the main focus of this paper is on the tidal deformability, we will focus on this observable in this appendix and compare the sGB correction to the leading tidal effect in the gravitational waveform from a binary neutron star inspiral.

There are two types of corrections to the waveform, (i) dissipative (gravitational wave luminosity) and (ii) conservative (Kepler's law). Since the scalar charges are zero for NSs in sGB gravity, the former correction vanishes to the post-Newtonian (PN) order\footnote{A term is said to be of order $n$PN if it is of $v^{2n} = (\pi m_t f)^{2n/3}$ relative to the leading where $v$ is the relative velocity of the binary constituents while $m_t = m_1+m_2$ is the total mass and $f$ is the gravitational wave frequency.} that has been computed to date~\cite{Shiralilou:2020gah,Shiralilou:2021mfl} (partially up to 3PN in our PN counting). On the other hand, Ref.~\cite{Tahura:2018zuq} showed that when the Kepler's law is corrected as
\begin{equation}
\Omega^2 = \frac{m_t}{r^3}\left[1+\frac{1}{2} A p \left(\frac{m_t}{r}\right)^p\right]\,,
\end{equation}
where $A$ and $p$ are some parameters characterizing the non-GR effect, the correction to the gravitational wave phase in the frequency domain is given by  
\begin{equation}
\delta \Psi_\mathrm{non-GR}=-\frac{5}{32} A \frac{2p^2-2p-3}{(4-p)(5-2p)}\eta^{-2p/5} u^{2p-5}\,,
\end{equation}
where $\eta = m_1 m_2/m_t^2$ is the symmetric mass ratio while $u = (\pi \mathcal{M} f)^{1/3}$ for the chirp mass $\mathcal{M}= m_t \eta^{3/5}$. For sGB gravity, $A = (128/3)\zeta$ and $p=6$~\cite{Saffer:2019hqn}, so 
\begin{equation}
\delta \Psi_\mathrm{sGB}=-\frac{190}{7} \zeta\eta^{-12/5} u^{7}\,.
\end{equation}
In particular, for an equal-mass binary, this becomes
\begin{equation}
\label{eq:Psi_sGB}
\delta \Psi_\mathrm{sGB}= - \frac{760}{7} \zeta\left( \pi m_t f\right)^{7/3}\,.
\end{equation}
Given that the GR leading term is proportional to $(\pi m_t f)^{-5/3}$, the above correction is a 6PN effect.

\begin{figure}
    \centering
    \includegraphics[width=\linewidth]{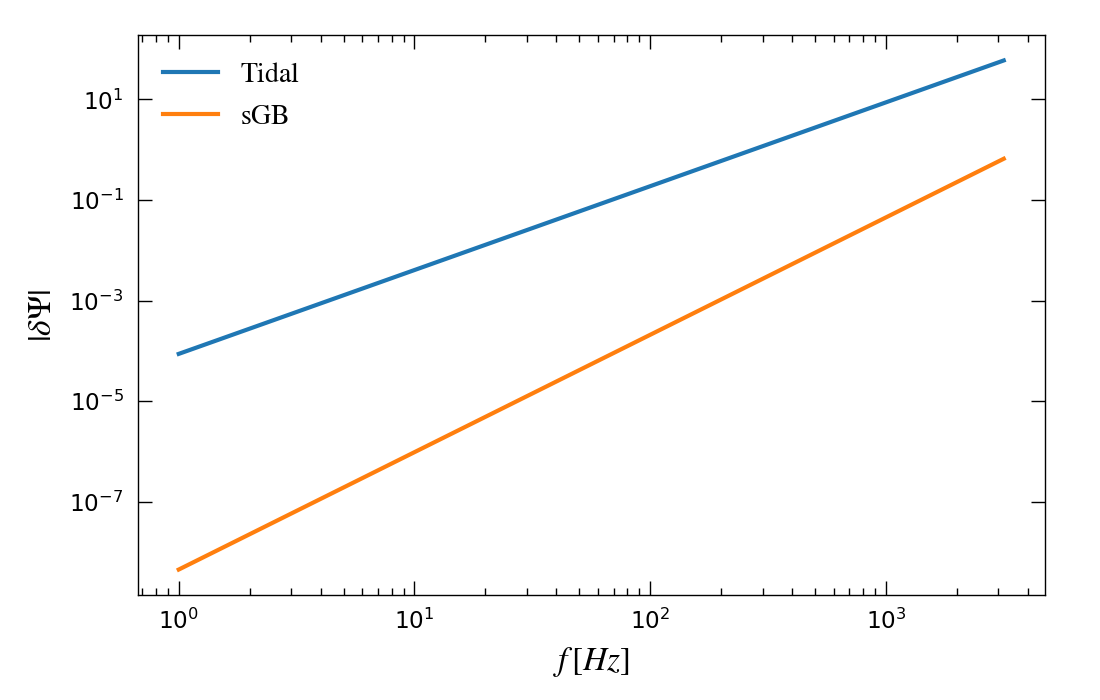}
    \caption{Comparison of the variation of the phase perturbation stemming from the leading tidal effect at 5PN order in GR  (Eq.~\eqref{eq:Psi_tidal}, blue) and the correction from sGB gravity (Eq.~\eqref{eq:Psi_sGB}, orange) for a binary system of two $1.4 M_\odot$ NSs constructed with the MPA1 EoS.  $\zeta = 0.63$ is the largest correction availible for MPA1 as seen in Fig.~\ref{fig:MRcurve_max_mass}. Notice the sGB correction is much smaller than the GR tidal contribution and is thus negligible.}
    \label{fig:SystematicComparison}
\end{figure}

Let us next compare the above with the leading tidal effect in the GR waveform that enters at 5PN order~\cite{Flanagan:2007ix,Yagi:2013awa}. For an equal-mass binary, it is given by
\begin{equation}
\label{eq:Psi_tidal}
\delta \Psi_\mathrm{tidal} = 
- \frac{117}{64} \Lambda \left( \pi m_t f\right)^{5/3}\,.
\end{equation}
Figure~\ref{fig:SystematicComparison} compares Eqs.~\eqref{eq:Psi_sGB} and~\eqref{eq:Psi_tidal} as a function of the gravitational wave frequency for a selected mass, EoS and $\zeta$. Notice that the sGB correction is always suppressed (at least by three orders of magnitude) than the GR tidal effect. This is because the former enters at higher PN order and the latter is enhanced by $\Lambda$ which is $\sim 895$ for the example system in Fig.~\ref{fig:SystematicComparison}. This suggests that the sGB correction can only affect the measurement of the tidal deformability by $\sim 0.1\%$ at most and thus negligible.

\section{Constant Density Star}
\label{app:CD}

In this section, we present an analytic result for the sGB correction to the Love number and tidal deformability. In order to have the calculations analytically tractable, we focus on constant density  (or incompressible) stars and work in the Newtonian limit. This amounts to keeping only the leading, non-vanishing contribution within the post-Minkowskian approximation (expansion in small $m/r$ or $M/R$) at each order in $\order{\epsilon^n,\alpha^m}$. In this appendix, we set $\rho_{00} = \rho_c$ in the interior region.

\subsection{Metric at $\order{\epsilon^0,\alpha^0}$}

Let us first study the GR background. By taking the Newtonian limit of Eqs.~\eqref{eq:m0_eq}--\eqref{eq:pressure_gr}, we find
\begin{subequations}
\label{eq:GR_n0}
    \begin{align}
    \diff{m}&= 4 \pi \rho_{c} r^2\,, \\
    \diff{\tau_{00}} &= \frac{2 m }{r^2}\,, \\
    \sigma_{00} &= \frac{2 m }{r}\,, \\
    \diff{p_{00}} &= \frac{\rho_c m}{r^2}\,,
    \end{align}
\end{subequations}
in the interior region. The above equations can easily be solved as
\begin{subequations}
\begin{align}
m^\mathrm{int} &= \frac{4}{3}\pi\rho_c r^3\,, \\
\label{eq:tau00_const}
\tau_{00}^\mathrm{int} &= \frac{4}{3} \pi \rho_c \left(r^2-3 R_0^2\right)\,, \\
p_{00}^\mathrm{int} &= p_c+\frac{2}{3} \pi  \rho_c^2 r^2\,.
\end{align}
\end{subequations}
In the exterior region, one can set $\rho_c=0$ and $p_{00} = 0$ to yield
\begin{equation}
m^\mathrm{ext} = M_0\,, \quad
\tau_{00}^\mathrm{ext} = -2\frac{M_0}{r}\,.
\end{equation}
The integration constant in Eq.~\eqref{eq:tau00_const} has been determined by matching the interior and exterior solutions at the surface $r=R_0$.

\subsection{Metric at $\order{\epsilon^1,\alpha^0}$}

Next, we look at the tidal perturbation for constant density stars in GR following~\cite{Damour:2009vw}. First, the Newtonian limit of Eq.~\eqref{eq:HindererEquation} is given by~\cite{Hinderer:2007mb}
\begin{equation}
\label{eq:tau10_const}
\frac{d^2 \, \tau_{10}^\mathrm{int}}{dr^2} + \frac{2}{r} \frac{d\,\tau_{10}^\mathrm{int}}{dr} - \left(\frac{6}{r^2} + \frac{3}{\rho_c r} \frac{d\,\rho_{c}}{dr}\right)\tau_{10}^\mathrm{int}=0\,,
\end{equation}
while $K_{10} = - \tau_{10}$ to leading order.
When finding solutions in the interior and exterior regions up to integration constants, we can set the last term in the above equation to 0, since the $d\rho_c/dr$ term will only contribute to the boundary condition which will cause a discontinuity at the surface (we will deal with this below).  Then, we find
\begin{subequations}
\begin{align}
\label{eq:tau_int_const}
\tau_{10}^\mathrm{int} &= a_0 r^2\,, \\ 
\label{eq:tau_ext_const}
\tau_{10}^\mathrm{ext} &= \frac{8}{5}c_1 \left(\frac{M_0}{r}\right)^3 + 3 c_2 \left( \frac{r}{M_0}\right)^2\,,
\end{align}
\end{subequations}
for the interior (regular at the center) and exterior solutions, respectively. Notice that these solutions are equivalent to Eqs.~\eqref{eq:tau10InteriorExpansion} and~\eqref{eq:tau10ExteriorExpansion} but removing ``$+\mathcal{O}(r^n)$''.

We next discuss the boundary conditions at the surface.
Notice that this equation contains a derivative of the density.  Since the density is discontinuous at the boundary, we must be careful when matching the interior and exterior solutions.
Namely, if we redefine our density as
\begin{equation}
\rho_{00} = \rho_c \, \Theta(R_0-r)\,,
\end{equation}
with a Heaviside function $\Theta$,
we see that this last term does affect the final result at the boundary of the star.  This leads to a singular term and we have two new equations to solve for at the boundary:
\begin{subequations}
\label{eq:BoundaryConditions_n0}
    \begin{align}
    \tau_{10}^{\rm int} (R_0) &= \tau_{10}^{\rm ext} (R_0) \,, \\
    \frac{R_0}{\tau_{10}^{\rm int}} \frac{d\,\tau_{10}^{\rm int}}{dr}\bigg|_{R_0} -3 &= \frac{R_0}{\tau_{10}^{\rm ext}} \frac{d\,\tau_{10}^{\rm ext}}{dr}\bigg|_{R_0}\,.
    \end{align}
\end{subequations}
We use Eqs.~\eqref{eq:tau_int_const} and~\eqref{eq:tau_ext_const} and solve the above boundary conditions for $c_1$ and $c_2$ to yield
\begin{equation}
\label{eq:c_output}
    c_1 = \frac{3 R_0^5}{8M_0^3} a_0\,,\quad
    c_2 = \frac{2M_0^2}{15} a_0\,.
\end{equation}
From this, we find the Love number $k_2$ to be a constant $k_2=0.75$~\cite{1955MNRAS.115..101B}, or equivalently $\Lambda_0 = 1/\left(2\mathcal{C}_0^5\right)$
~\cite{Yagi:2013awa}, where $\mathcal{C}_0$ is the compactness of the star.

\subsection{Scalar Field at $\order{\epsilon^0,\alpha^1}$}

We now turn our attention to the background scalar field. In the Newtonian limit for constant density stars, the field equation is given by
\begin{equation}
\label{eq:phi01_int}
\frac{d^2\varphi_{01}}{dr^2} + \frac{2}{r}\frac{d\varphi_{01}}{dr}={
\frac {256\,{\pi}^{2}{\rho_{c}}^{2}}{3}}\,,
\end{equation}
in the interior while the source term on the right hand side is absent in the exterior.
Solving this equation in the interior and exterior regions with regularity at the center and infinity, we find
\begin{subequations}
    \begin{align}
        \varphi^{\rm int}_{01} &= \frac{128 \pi}{9} \rho_c^2 r^3 + \varphi^{(c)}_{01} \,, \\
        \varphi^{\rm ext}_{01} &= - \frac{4 M_0^2}{r^4}\,.
    \end{align}
\end{subequations}
Here we have set the constant term in the exterior region to 0. The integration constant $\varphi^{(c)}_{01}$ in the interior solution can be determined through the matching of the two solutions at the boundary, though it does not affect the calculations below as the scalar field only enters through its derivatives in the field equations.

\subsection{Metric at $\order{\epsilon^0,\alpha^2}$}

We now comment on the sGB metric and matter corrections at the background level. First we set $\rho_2=0$. This is because 
$\rho_2$ is a free parameter for constant density stars and we simply use $\rho_c$ as the value of the full central density for constant density stars in sGB gravity. Then, we find that the source terms for the differential equations for $\tau_{02}$, $\sigma_{02}$ and $p_2$ enter at $\order M^3$,  $\order M^3$, and  $\order M^4$, respectively in terms of the post-Minkowskian order counting. This means that the solutions enter at the same orders, which only give higher order corrections to the tidal Love numbers. Thus, we can safely ignore the contribution at $\order{\epsilon^0,\alpha^2}$ in the following analysis.

\subsection{Scalar Field at $\order{\epsilon^1,\alpha^1}$}

We not look at the tidal perturbation to the scalar field. Keeping only the leading source term within the post-Minkowskian analysis, the field equation in the interior region is given by
\begin{equation}
\label{eq:phi11}
\frac{d^2 \varphi_{11}}{dr^2} + \frac{2}{r} \frac{d\varphi_{11}}{dr} - \frac{6}{r^2}\varphi_{11} = S_{\varphi_{11}}\,,
\end{equation}
with
\begin{equation}
S_{\varphi_{11}}^\mathrm{int} = 64\,\pi \,a_{0} r \frac{d\rho_c}{dr}\,, \quad S_{\varphi_{11}}^\mathrm{ext} = {\frac {144 c_{2}}{M_0{r}^{3}}}\,,
\end{equation}
for the interior and exterior sources, respectively.
Compared to Eq.~\eqref{eq:phi01_int}, the third term on the left hand side in Eq.~\eqref{eq:phi11} is due to the fact that we are looking at the quadrupolar tidal perturbation. The above equation can be solved under the boundary condition of regularity at the center and infinity to yield
\begin{equation}
\varphi^{\rm int}_{11} = \varphi^{(c)}_{11} r^2 \,, \quad
\varphi^{\rm ext}_{11} = \frac{\phi_5}{r^3} - \frac{24 c_2}{M_0 r}\,.
\end{equation}

The integration constants can be determined from the bounary condition at the surface. Similar to the case at $\order{\epsilon^1,\alpha^0}$, there is a term proportional to $d\rho_c/dr$ in Eq.~\eqref{eq:phi11} that becomes singular at the surface and contributes to the boundary condition as
\begin{subequations}
\begin{align}
\varphi^{\rm int}_{11}(R_0) &= \varphi^{\rm ext}_{11}(R_0)\,, \\
\frac{d\varphi^{\rm int}_{11}}{dr}\bigg|_{R_0}  -64 R_0 \pi a_0 \rho_c &= \frac{d\varphi^{\rm ext}_{11}}{dr}\bigg|_{R_0}\,.
\end{align}
\end{subequations}
Using these, we find the solution for $\varphi_{11}$ as
\begin{equation}
\varphi^{\rm int}_{11} = \frac {208\,a_{0}\,M_0}{25\,{R_0}^{3}}{r}^{2}\,, \quad \varphi^{\rm ext}_{11} = {\frac {16\,a_{0}\,M_0 \left( 18\,{R}^{2}-5\,{r}^{2} \right) }{25\,{r}^{
3}}}\,.
\end{equation}

\subsection{Metric at $\order{\epsilon^1,\alpha^2}$}

The final step is the solution at $\order{\epsilon,\alpha^2}$.
Following the methodology laid out in Sec.~\ref{subsec:LoveCorrection}, one can derive an equation for $\tau_{12}$. The interior and exterior equations are given by
\begin{widetext}
\begin{subequations}
\begin{align}
\label{eq:tau12_const}
\frac{d^2 \, \tau_{12}^\mathrm{int}}{dr^2} + \frac{2}{r} \frac{d\,\tau_{12}^\mathrm{int}}{dr} - \left(\frac{6}{r^2} + \frac{3}{\rho_c r} \frac{d\,\rho_{c}}{dr}\right)\tau_{12}^\mathrm{int}&=16384\,{\pi}^{3} a_{0}\rho_{c}  \frac{d\rho_c}{dr}   r
\,, \\
\frac{d^2 \, \tau_{12}^\mathrm{ext}}{dr^2} + \frac{2}{r} \frac{d\,\tau_{12}^\mathrm{ext}}{dr} - \frac{6}{r^2} \tau_{12}^\mathrm{ext}&= {\frac {2048\,\pi}{25{R}_0^{3} \,{r}^{11}}} \left( 576\,{M}_0^{2}{R}_0^{5}a_{0}\,{r}^{3}+75\,{M}_0^{
2}{R}_0^{3}a_{0}\,{r}^{5}+52\,{M}_0^{2}a_{0}\,{r}^{8} \right. \nonumber \\
& \quad \left.  +1240\,{M}_0^{5}{R}_0^{3}
c_{1}-225\,{R}_0^{3}c_{2}\,{r}^{5} \right)\,,
\end{align}
\end{subequations}
\end{widetext}
respectively. We solve the above equations with regularity at the center and infinity to find
\begin{subequations}
    \begin{align}
    \tau_{12}^{\rm int} &= \tau_{12}^{\rm (c)} r^2 \,, \\
    \tau_{12}^{\rm ext} &= - \frac{53248\pi a_0 M_0^2}{75 R_0^3 \, r} + \frac{\tau_3}{r^3} + \frac{3072 \pi a_0 M_0^2}{5\, r^ 4} \nonumber \\
    & \quad + \frac{49152 \pi a_0 R_0^2 M_0^2}{25 r^6} + \frac{31744 \pi a_0 R_0^5 M_0^2}{55 \, r^9}\,.
    \end{align}
\end{subequations}
with integration constants $\tau_{12}^\mathrm{(c)}$ and $\tau_3$, which are determined from the boundary condition at the surface. Taking into account singular contribution from $d\rho_c/dr$ in Eq.~\eqref{eq:tau12_const}, the boundary condition is given by
\begin{subequations}
\begin{align}
\tau_{12}^{\rm int}(R_0) &= \tau_{12}^{\rm ext}(R_0)\,, \\
\frac{d\tau_{12}^{\rm int}}{dr}\bigg|_{R_0}  -R_0\left( 16384 \pi^3 a_0 \rho_c^2 + 3 \tau_{12}^{\rm (c)}\right) &= \frac{d\tau_{12}^{\rm ext}}{dr}\bigg|_{R_0}\,.
\end{align}
\end{subequations}
From these, we find the integration constants to be 
\begin{subequations}
    \begin{align}
    \tau_{12}^{\rm (c)} &=  - \frac{899072 \pi a_0 M_0^2}{825 R_0^6}\,, \\
    \tau_{3} &= - \frac{38912 \pi a_0 M_0^2}{11 R_0}\,.
    \end{align}
\end{subequations}

\begin{figure}
	\includegraphics[width=\linewidth]{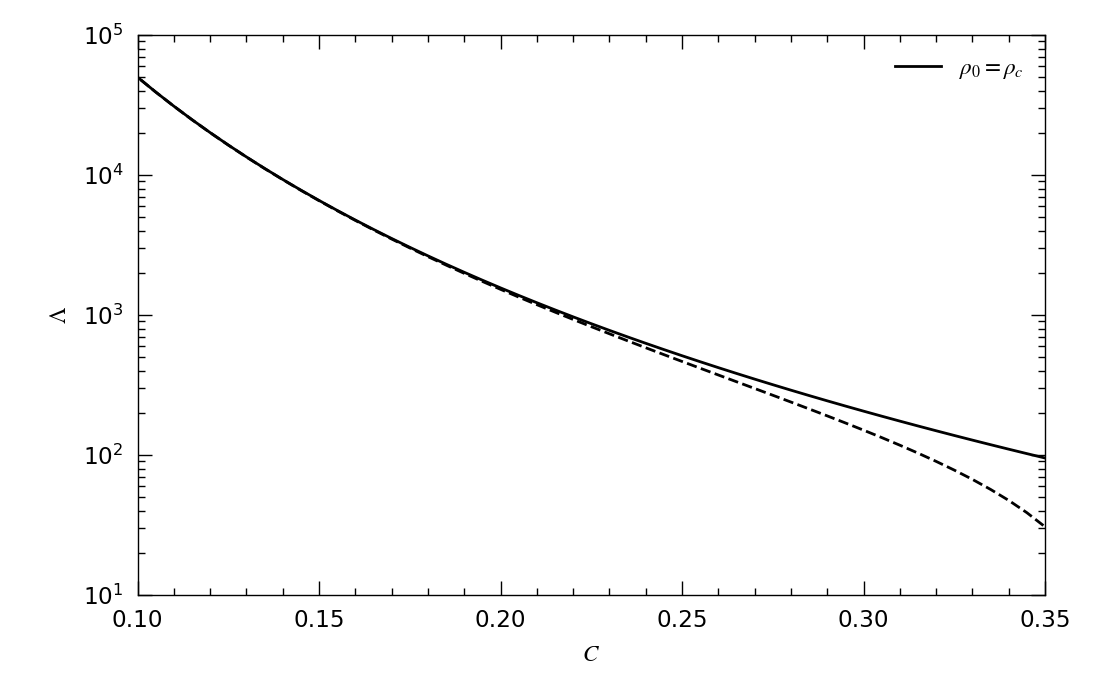}
    \caption{
$\Lambda$--$C$ relation for constant density stars in GR (solid) and sGB gravity with $\zeta=1$ (dashed) in the Newtonian limit.
}
    \label{fig:ConstRho}
\end{figure}

With our metric now fully solved for (to our leading non-trivial Newtonian order), we may continue with the procedure presented in Sec.~\ref{sec:LoveAndTidal}.  These lead us to have modifications to the Love number of the form
\begin{subequations}
    \begin{align}
    k_2 &=  \frac{3}{4} - \frac{3040}{11} C_0^6 \zeta \,, \\
    \Lambda &= \frac{1}{2 C_0^5} - \frac{6080}{33} C_0 \zeta \,.
    \end{align}
\end{subequations}
Figure~\ref{fig:ConstRho} shows the results for the modification to the $\Lambda-C$ relation for a constant density star to leading Newtonian order. Notice that the qualitative feature is similar to that for realistic NSs in Fig.~\ref{fig:FullLambdaC}.

\bibliography{bibliography.bib}

\end{document}